\documentclass[fleqn,usenatbib]{mnras}
\usepackage{graphicx}	% Including figure files
\usepackage{amsmath}	% Advanced maths commands
\usepackage{amssymb}	% Extra maths symbols

\graphicspath{{./}}

\title[IMF in the low-metallicity environment]
{Transition of the initial mass function in the metal-poor environments}

\author[S. Chon, K. Omukai, \& R. Schneider]{
Sunmyon Chon $^{1}$\thanks{E-mail: sunmyon.chon@astr.tohoku.ac.jp},
Kazuyuki Omukai $^{1}$, and
Raffaella Schneider $^{2,3,4}$
\\
$^{1}$Astronomical Institute, Graduate School of Science, Tohoku University, Aoba, Sendai 980-8578, Japan\\
$^{2}$Dipartimento di Fisica, Universit\`{a} di Roma ‘La Sapienza’, P.le Aldo Moro 2, I-00185 Roma, Italy \\
$^{3}$INAF/Osservatorio Astronomico di Roma, via di Frascati 33, I-00078 Monteporzio Catone, Italy \\
$^{4}$INFN, Sezione di Roma 1, P.le Aldo Moro 2, I-00185 Roma, Italy
}

\date{Accepted 2021. Received 2021; in original form 2021}

\pubyear{2021}

\begin{document}

\label{firstpage}
\pagerange{\pageref{firstpage}--\pageref{lastpage}}
\maketitle

\graphicspath{./}

\begin{abstract}
We study star cluster formation in a low-metallicity environment 
using three dimensional hydrodynamic simulations.
Starting from a turbulent cloud core, 
we follow the formation and growth of protostellar systems
with different metallicities ranging from $10^{-6}$ to $0.1~Z_{\odot}$.
The cooling induced by dust grains promotes 
fragmentation at small scales and 
the formation of low-mass stars with $M_{*} \sim 0.01$--$0.1~M_{\odot}$
when $Z/Z_{\odot} \gtrsim 10^{-5}$.
While the number of low-mass stars increases with metallicity,
the stellar mass distribution is still top-heavy for $Z/Z_{\odot} \lesssim 10^{-2}$
compared to the Chabrier initial mass function (IMF).
In these cases, star formation begins after the turbulent motion decays
and a single massive cloud core monolithically collapses
to form a central massive stellar system.
The circumstellar disk preferentially feeds 
the mass to the central massive stars,
making the mass distribution top-heavy.
When $Z/Z_{\odot}=0.1$, 
collisions of the turbulent flows promote the onset of the star formation
and a highly filamentary structure develops
owing to efficient fine-structure line cooling.
In this case, the mass supply to the massive stars is limited by the local gas reservoir
and the mass is shared among the stars,
leading to a Chabrier-like IMF.
We conclude that cooling at the scales of the turbulent motion promotes the development of the filamentary structure
and works as an important factor leading to the present-day IMF.
\end{abstract}


\section{Introduction}
The initial mass function (IMF) plays a crucial role in 
the formation and evolution of the stars and galaxies that we observe today.
In the primordial and low-metallicity environments
the nature of the IMF is still poorly understood, 
which limits our knowledge about early structure formation.
Among a spectrum of stellar masses, 
massive stars dramatically change the surrounding environment:
they scatter the heavy elements around when they explode as supernovae 
\citep{Wise+2012,Ritter+2015, Graziani+2015, Graziani+2017, Sluder+2016, Bennassuti+2017,Chiaki+2018}
and emit a large amount of ionizing photons that ionize the surrounding gas 
\citep{Johnson+2013,Chon+2017,Nakatani+2020}.
The extremely massive stars ($40 < M_{*}/M_{\odot} < 120$ or $260 < M_{*}/M_{\odot}$)
directly collapse into the black holes \citep[BHs, e.g.][]{HegerWoosley2002},
providing a potential formation channel to the seeds of the observed supermassive BHs 
\citep[e.g.][]{Hosokawa+2012, Latif+2013, Chon+2016, Valiante+2016, Regan+2017, Wise+2019, Inayoshi+2020} 
and also contributing to cosmic reionization 
as additional UV radiation is emitted when the mass falls onto the BHs 
\citep[e.g.][]{Johnson+2007,Alvarez+2009,Johnson+2011,Jeon+2012,Aykutalp+2013,Aykutalp+2014,Chon+2021}.

Recent numerical simulations have revealed that 
the first stars or population III (Pop~III) stars are typically massive,
with masses in the range $10$--$100~M_{\odot}$ 
\citep{OmukaiNishi1998,Bromm+2001, OmukaiPalla2003, Yoshida+2008, Hosokawa+2011, Hirano+2014, Hirano+2015, Susa+2014, Hosokawa+2016, Sugimura+2020},
accompanied by a small number of low-mass stars \citep{Machida+2008, Clark+2011, Greif+2012, Machida&Doi2013, Stacy+2016, Susa2019}.
The Pop~III mass spectrum at the lower-mass end is 
constrained by the number density of metal-poor stars observed in Galactic halo and local dwarf galaxies,
indicating a top-heavy IMF in the primordial environments
\citep{Salvadori+2007, Salvadori+2008,Bennassuti+2014, Bennassuti+2017, Graziani+2015, Ishiyama+2016, Hartwig+2018b, Magg+2018}.
In contrast, observational studies of the IMF in the present-day universe have shown that
present-day stars typically have small masses
\citep[e.g.][]{Salpeter1955,Kroupa2002,Chabrier2003}.
The observationally derived IMF has a universal distribution,
showing little environmental dependence, 
and is peaked at $0.1~M_{\odot}$.
Still, what drives such transitions, from the top-heavy to the present-day IMF,
remains unresolved today.

Thermal processes inside a cloud can determine 
the typical fragmentation scales of the collapsing cloud
and thus give a typical mass of the stars
\citep{Larson1985, Larson2005, Bonnell+2006}.
The cloud becomes highly filamentary and starts to fragment
when cooling is efficient and 
the effective specific heat ratio $\gamma$ is smaller than unity, where
$\gamma \equiv \mathrm{d} \log P / \mathrm{d} \log \rho$ and
$P$ and $\rho$ are gas pressure and density, respectively
\citep[e.g.][]{Inutsuka&Miyama1992,Li+2003,Jappsen+2005}.
Once $\gamma$ becomes larger than unity, 
the growth of such filamentary modes ceases
and further fragmentation is suppressed,
setting a minimum scale for fragmentation.

The presence of metals significantly alters the thermal evolution of the cloud,
reducing the gas temperature and the characteristic mass of the stars.
The abundance of metals in the gas phase and locked in dust grains
is thus thought to be one of the key parameters 
which drives the IMF transition 
\citep[e.g.][]{Omukai2000,Schneider+2003,Omukai+2005, Schneider+2006, Schneider&Omukai2010, Schneider+2012, Chiaki+2014}.
In recent three-dimensional simulations, cloud fragmentation is observed
when the cloud is enriched to a finite metallicity
\citep{Clark+2008, Dopcke+2011, Dopcke+2013, Safranek-Shrader+2016, Chiaki+2016, Chiaki+2020}.
They find that a number of low-mass stars with $M_{*} \lesssim 0.01$--$0.1~M_\odot$ form
due to the fragmentation induced by the efficient cooling by dust grains.
Even a trace amount of dust 
with the dust-to-gas mass ratio of $10^{-9}$ -- $10^{-8}$
(which is $10^{-6}$ times the solar neighborhood value)
can trigger the formation of low-mass stars
\citep{Schneider+2012}.

So far, the stellar mass distribution derived for low-metallicity environments
cannot be interpreted as the IMF.
Most of the previous studies do not follow the stellar mass evolution
for a sufficiently long time-scale, $10^{4}$ -- $10^{5}~$years,
nor do not adequately resolve the spatial scales for fragmentation.
Since stars are still accreting gas at the end of the simulations,
there is still the possibility that massive stars efficiently grows more massive, 
leading to a top-heavy IMF.
For example, \citet{Dopcke+2013} followed the evolution only for $120~$yrs from the formation of the first protostar, which is not enough to discuss the final IMF shape. 
Star cluster formation in the present-day universe suggests that
massive stars grow in mass 
while the continuous formation of low-mass stars
keeps the shape of the mass function unchanged
\citep[e.g.][]{Bate+2003}.

Another difficult issue for star formation in low-metallicity environments is
that the initial conditions are poorly constrained.
Along with cosmic structure formation,
not only the metallicity but other important factors evolve with time.
The evolution of the abundance pattern of metals or the composition of dust grains
\citep{Schneider+2012, Chiaki+2016, Chiaki+2020},
heating by the warmer cosmic microwave background (CMB) at high redshift
\citep{Smith+2007, Jappsen+2009, Schneider&Omukai2010, Meece+2014, Bovino+2014, Safranek-Shrader+2014, Riaz+2020},
the strength of the turbulent motion induced by the expansion of H~{\sc ii}  regions
or by SN explosions \citep{Smith+2015, Chiaki+2018}, and
the effects of the magnetic fields generated at some cosmic evolutionary stages
\citep{Machida+2009,Machida&Nakamura2015,Higuchi+2018}.
While these points are important to fully understand the star formation process 
during the evolution of cosmic structures,
they are tangled up making it difficult to capture the metallicity effect
on the stellar mass distribution.

In this paper, we perform high resolution hydrodynamics simulations following the gravitational collapse of a turbulent cloud core simultaneously solving non-equilibrium chemical reaction network and the associated cooling processes.
In order to capture the mass spectra in low-metallicity environments,
the simulations resolve the protostellar $\sim$au scale
and extend for $10^{4}$--$10^{5}~$years,
which is a typical time-scale for star formation.
We perform the runs assuming six different metallicity values
with $10^{-6} \lesssim Z/Z_{\odot} \lesssim 0.1$,
starting from the same initial conditions.
This allows us to evaluate impact of the metallicity 
on the resulting stellar mass distribution. 
This is the first systematic attempt to investigate how the IMF evolves from the primordial to present-day environments by means of detailed numerical simulations. 

This paper is organized as follows.
In Section~\ref{sec::methodology}, 
we describe the initial conditions and our numerical methodology.
Section~\ref{sec::results} presents our numerical results.
We discuss the implication of our results and some caveats in Section~\ref{sec::discussion}.
Concluding remarks are given in Section~\ref{sec::summary}.

\section{Methodology} 
\label{sec::methodology}
We perform hydrodynamics simulations
using the smoothed particle hydrodynamic (SPH) code, {\tt Gadget-2} \citep{Springel2005}.
Starting from an initial turbulent cloud core,
we follow the cloud evolution, 
from the onset of cloud collapse 
to the formation of numerous protostars.
Deriving the protostellar mass functions for different cloud metallicities,
we focus on how the cloud metallicity modifies the cloud evolution and 
thus the mass function of the stars.
Here, we describe the initial conditions and 
overview the numerical methodology 
additionally implemented into the original version of {\tt Gadget-2} code.

As the initial conditions, we generate the gravitationally unstable critical Bonnor-Ebert sphere, which is characterized by the central gas density $n_\text{c}$ and temperature $T_\text{c}$.
We take $n_\text{c}=10^{4}~\mathrm{cm^{-3}}$ and $T_\text{c} = 200~$K,
which sets the cloud radius to be $1.85~$pc.
We then enhance the cloud density by a factor of $1.4$ to promote gravitational collapse of the cloud,
which results in a total cloud mass of $1950~M_{\odot}$.
We employ $4.04 \times 10^{6}$ SPH particles to construct the initial gas sphere,
where the particle mass is $4.8\times10^{-4}~M_{\odot}$.

We also include rotational and turbulent motions to model the initial velocity field.
The turbulent velocity field is generated following \citet{MacLow1999}.
We divide the whole simulated region using a $128^{3}$ grid and 
assign a random velocity component to each grid cell,
assuming it follows a random Gaussian fluctuation with a power-law power spectrum, $P(k) \propto k^{-4}$.
We then provide each gas particle with a turbulent velocity,
interpolating it from the values in the surrounding grid cells (e.g. cloud-in-cell interpolation).
We impose transonic turbulence with $M_\text{ch} \equiv v_\text{rms}/c_\text{s} = 1$,
where $M_\text{ch}$ is the Mach number, $v_\text{rms}$ is the root mean square of the turbulent velocity, 
and $c_\text{s}$ is the sound speed of the gas with temperature $T_\text{c}$.
This gives $v_\text{rms} = 1.17~\mathrm{km~s^{-1}}$,
similar to the turbulent velocity observed in local molecular clouds at pc scale. \citep{Larson1981,Heyer+2004}.
We also impose rigid rotation on the $x-y$ plane
so that the rotational energy equals 0.1\% of the total gravitational energy,
which roughly corresponds to the observed rotational energy of the molecular cloud cores, which is in the range $10^{-4} - 0.07$ \citep{Caselli+2002}.

During the gravitational collapse of the cloud core,
we must be capable of resolving the local Jeans mass $M_\text{J}$ with more than $100$--$1000$ particles
\citep[e.g.][]{Bate+1995, Truelove1997, Stacy+2013}.
For this purpose, once the gas density exceeds $10^{6}~\mathrm{cm^{-3}}$,
we split each gas particle into $13$ daughter particles
following the method presented in \citet{Kitsionas+2002}.
The mass of each new particle is thus $3.7\times 10^{-5}~M_{\odot}$,
leading to the minimum resolvable mass scale 
$M_\text{res} \sim 1.5N_\text{neigh} m_\text{SPH} = 3.6 \times 10^{-3}~M_\odot$,
where $N_\text{neigh}=64$ is the number of the neighbour particles and
$m_\text{SPH}$ is the mass of an SPH particle \citep{BB1997}.
This allows us to resolve the Jeans mass of cloud cores throughout our simulation,
which has $0.01~M_\odot$ at its minimum \citep[e.g.][]{Bate2009, Bate2019}.

Once the gas density exceeds the critical value $n_\text{sink}$,
we introduce a sink particle,
assuming that a protostar forms inside the cloud core.
We set $n_\text{sink} = 2.5 \times 10^{16}~\mathrm{cm^{-3}}$
for the following reason.
Below this density, the cloud is in hydrostatic equilibrium
since the cloud is optically thick and supported by the pressure.
Once the cloud density exceeds $n_\text{sink}$, 
owing to the evaporation of dust grains and 
the chemical cooling caused by H$_{2}$ destruction,
the cloud starts to collapse onto the protostellar core,
which is similar to the second collapse 
observed in the solar metallicity case
\citep{Larson1969, Masunaga&Inutsuka2000}.
In reality, the gravitational collapse further continues 
until the density reaches $\sim 10^{20}~\mathrm{cm^{-3}}$.
Since resolving the formation of protostellar cores is computationally expensive, 
we introduce the sink particles and 
mask out the high density region around the protostars,
which allows us to follow the formation of the stellar clusters
for $\sim0.1~$million years.
We set the sink radius to be $1~$au. 
Once the gas particles enter the region inside the sink radius, they are assimilated to the sink particle.
We allow the merger of a sink particle pair
if their separation becomes smaller than the sum of the sink radii.

\subsection{Chemistry}
We solve non-equilibrium chemical network for 8 species 
(e$^{-}$, H, H$^{+}$, H$_{2}$, H$^{-}$, D, D$^{+}$, and HD)
and solve the associated cooling/heating processes.
The line cooling rates due to 
ro-vibrational transition of H$_{2}$ and HD molecules are given by
\begin{align}
\Lambda_\text{line} = \beta_\text{esc} \Lambda_\text{thin} e^{-\tau},
\end{align}
where $\beta_\text{esc}$ is the escape probability of the line emission,
$\Lambda_\text{thin}$ is the line cooling rate in the optically-thin case,
and $\tau$ is the optical depth of the continuum radiation.
We adopt fitted values of $\Lambda_\text{thin}$
given by \citet{Glover2015} for H$_{2}$ and by \citet{Lipovka+2005} for HD
and $\beta_\text{esc}$ by \citet{Fukushima+2018}.
We consider continuum cooling owing to the primordial species 
for the following six processes,
H free-bound, H free-free, H$^{-}$ free-free, H$^{-}$ free-bound, and
collision induced emission (CIE) of H$_{2}$-H$_{2}$ and H$_{2}$-He
\citep{Mayer+2005, Matsukoba+2019}.
We also include chemical heating/cooling associated 
with the formation/destruction of H$_{2}$.

\begin{figure*}
	\centering
		\includegraphics[width=16.cm]{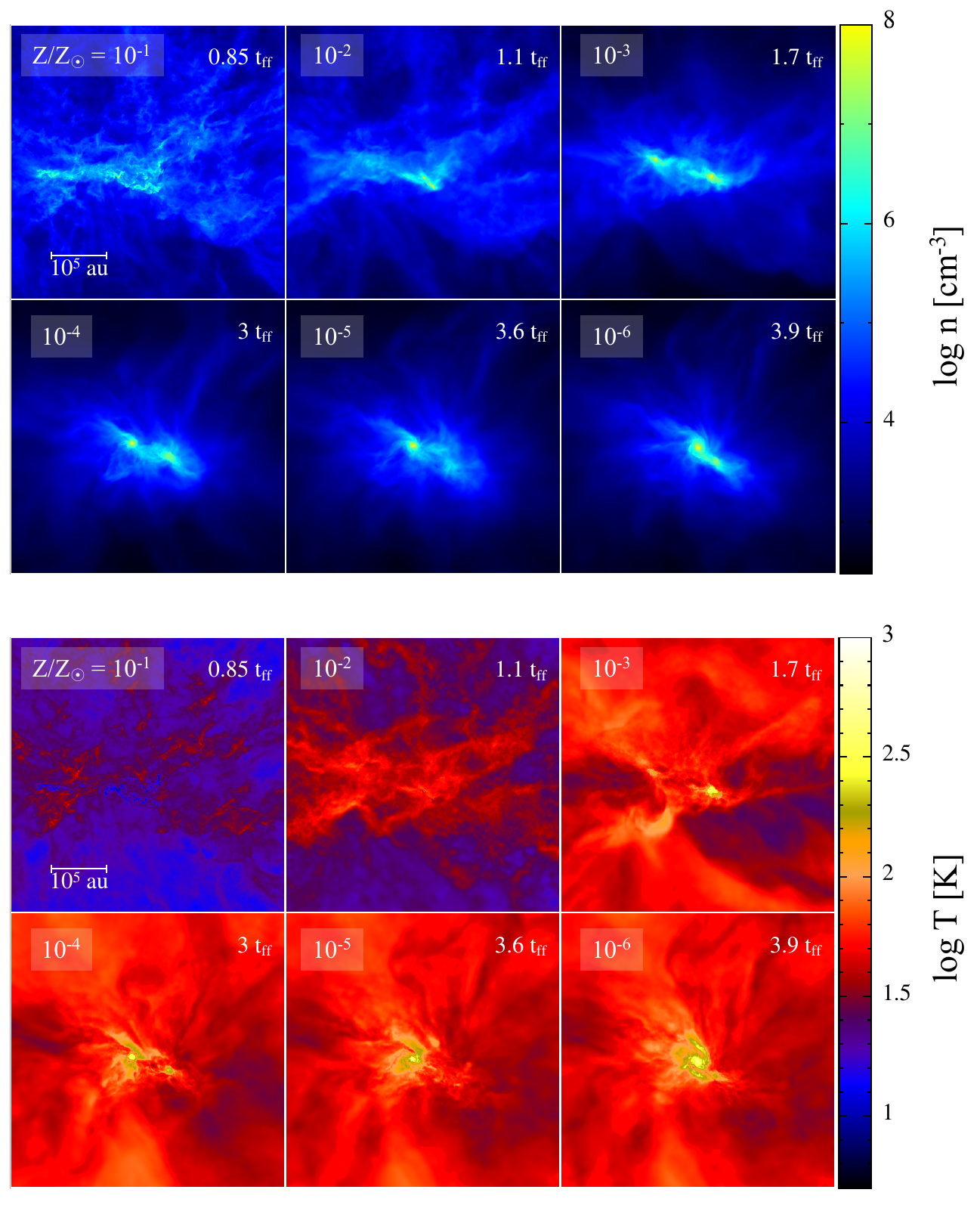}
		\caption{The projected density (top) and temperature distribution (bottom) along the $z$-axis 
		at the moment when the first protostar forms for the different metallicity cases. 
		We show the corresponding time of the snapshots in the right-top corner of each panel, 
		where $t_\text{ff}$ is the free-fall time of the initial cloud core.
		}
		\label{fig_snapshot}
\end{figure*}

To model cooling associated with heavy elements, 
we consider the atomic line transitions of C~{\sc ii}  and O~{\sc i}  and dust thermal emission.
We assume that the abundance ratio of heavy elements and
their fraction in the gas and dust phases
follow those in the solar neighborhood.
The amount of metals and dust are linearly scaled 
with a given metallicity $Z/Z_{\odot}$.
We use the dust opacity model given by \citet{Semenov+2003},
where the composition and the size distribution of dust grains
are the same as those found in the solar neighborhood.
We compute the dust temperature $T_\text{gr}$ 
from the thermal balance between heating and cooling of the dust grains,
solving the following equation,
\begin{align}
4\sigma T_\text{gr}^{4} \kappa_\text{gr} = \Lambda_\text{gas, dust} + 4 \sigma T_\text{rad}^{4} \kappa_\text{gr},
\end{align}
where $\sigma$ is the Stefan-Boltzmann constant, 
$\kappa_\text{gr}$ is the opacity of the dust,
$T_\text{rad}$ is the radiation temperature 
which is set to be the CMB temperature at $z=0$ ($2.72~$K),
and $\Lambda_\text{gas, dust}$ is the energy transfer between the gas and the dust grains due to the collisions,
which is taken from \citet{HM1979}. 

Here, we assume all the C and O in the gas phase are in the form of C~{\sc ii}  and O~{\sc i} 
and calculate the cooling rate by the fine-structure lines.
While the formation of molecules, such as CO, OH, and H$_{2}$O,
changes the chemical composition and thus slightly modifies the thermal evolution 
at intermediate metallicity cases $Z/Z_{\odot} \sim 10^{-4}$
\citep{Omukai+2005,Chiaki+2016},
the overall temperature evolution can be adequately described by our simple treatment.
In fact, one-zone calculation by \citet{Omukai+2005} have shown that
this model approximately 
reproduces the thermal evolution of low-metallicity clouds solved with 
the full non-equilibrium chemical network of C and O bearing species,
within the error of less than $30$\% in temperature.

\begin{figure}
	\centering
		\includegraphics[width=7.5cm]{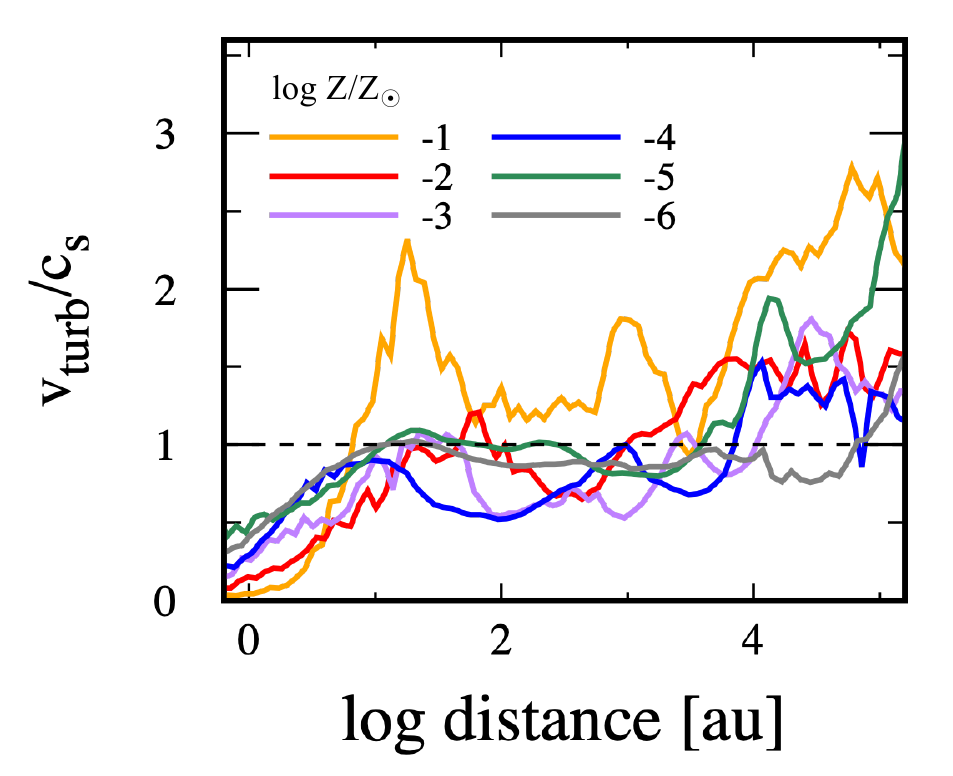}
		\caption{The radial profiles of the turbulent Mach number $v_\text{turb}/c_\text{s}$ for different metallicity values, with $\log Z/Z_\odot = -1$ (yellow), $-2$ (red), $-3$ (purple), $-4$ (blue), $-5$ (green), and $-6$ (grey line). The black dashed line represents the trans-sonic turbulence with $v_\text{turb}/c_\text{s} = 1$.
		}
		\label{fig_vturb}
\end{figure}

\begin{figure*}
	\centering
		\includegraphics[width=16.cm]{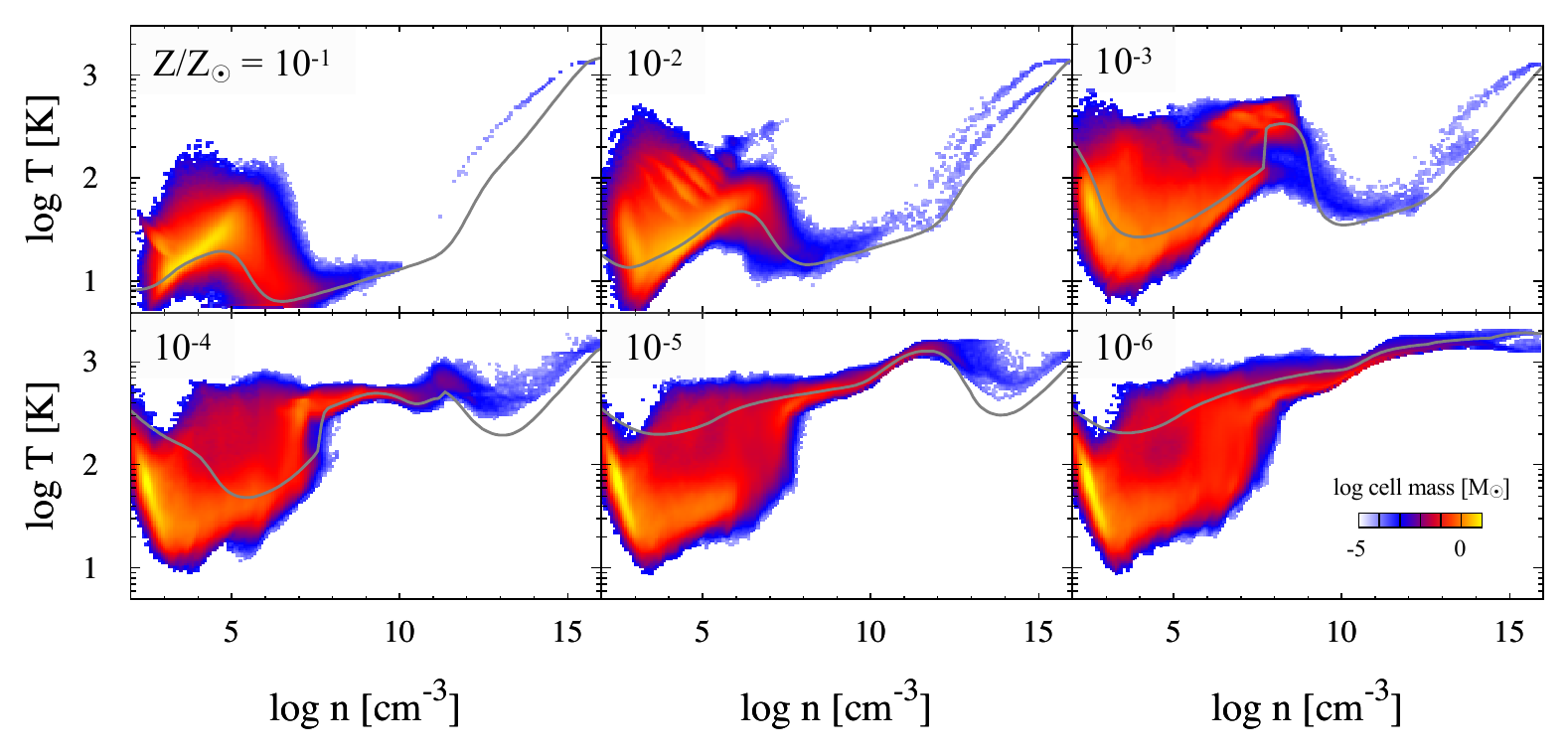}
		\caption{The temperature versus density diagram at the time of formation of the central protostar for the different metallicity models.
		Colors show the cell mass, where we divide the temperature and density region 
		by $200 \times 200$ cells and calculate the gas mass inside each cell.
		Solid grey lines show the temperature evolution obtained by one-zone calculation with the same metallicity.}
		\label{fig_rhoT_hist}
\end{figure*}

\begin{figure*}
	\centering
		\includegraphics[width=16.cm]{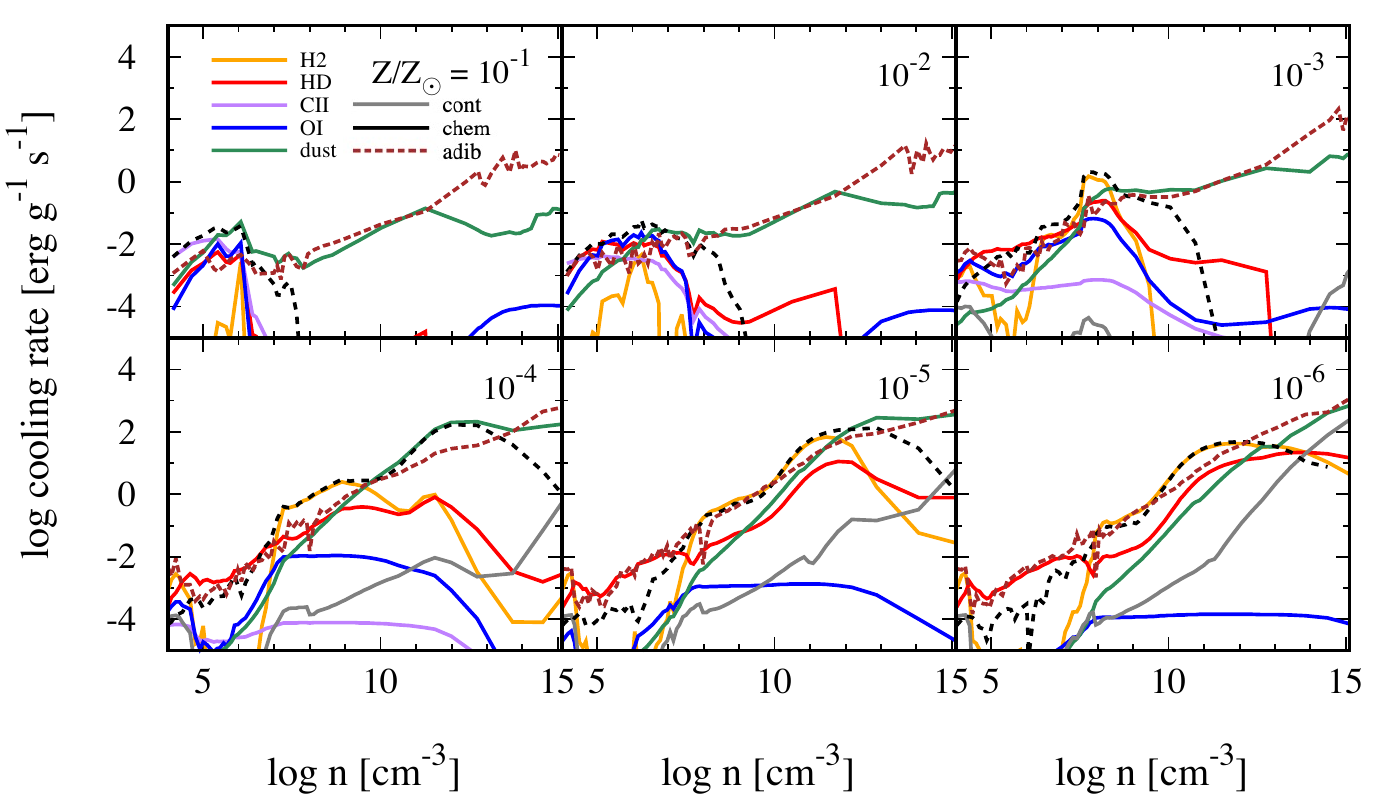}
		\caption{Cooling and heating rate at the cloud center
		as a function of the central gas density for different metallicity models.
		The lines with different colors show the cooling and heating rates 
		by H$_{2}$ (yellow), HD (red), C~{\sc ii}  (purple), O~{\sc i} (blue), dust thermal emission (green),
		continuum process (grey), chemical process (black), and adiabatic compression (brown).
		The solid and the dashed lines show the cooling and the heating rate, respectively.}
		\label{fig_cooling_rate}
\end{figure*}

%%%%%%%%%%%%%%%%%%%%%%%
%%%%%%%%%%%%%%%%%%%%%%%
\section{Results} 
\label{sec::results}
%%%%%%%%%%%%%%%%%%%%%%%
%%%%%%%%%%%%%%%%%%%%%%%
\subsection{Chemo-thermal evolution of the collapsing cloud}
We follow the collapse of a turbulent cloud core with different gas metallicities 
starting from the same initial conditions.
The top panel of Fig.~\ref{fig_snapshot} shows the density distribution 
at the moment when the first protostar is formed.
When $Z/Z_{\odot} = 10^{-1}$, filamentary structures develop due to the initial turbulent motion.
Such structures become less evident with decreasing metallicity.
When $Z/Z_{\odot} = 10^{-2}$, 
the density distribution at $n\lesssim 10^{5}~\mathrm{cm^{-3}}$
is similar to the case of $Z/Z_{\odot} = 10^{-1}$,
while the density fluctuation seeded by the turbulent motion has dissipated
at $n\gtrsim 10^{6}~\mathrm{cm^{-3}}$.
When $Z/Z_{\odot} \lesssim 10^{-4}$, 
the turbulent motion almost completely decays at the epoch of protostar formation.
In this case, a massive dense core forms at the cloud center,
in which the rotational motion dominates over the turbulence
to create a disk-like structure.

The key factor that distinguishes whether the filamentary structure develops 
is the efficiency of cooling in the shock compressed region.
If the turbulent motion becomes converging at some point, 
the gas density increases due to shock compression,
seeding a large density fluctuation.
When the shock compressed gas efficiently cools, 
it can condense further and 
the protostar forms once the gravitational instability operates therein.
The bottom panel of Fig.~\ref{fig_snapshot} shows the temperature distribution.
The cloud temperature decreases with increasing metallicity,
as the cooling becomes more efficient.
If we compare the temperature distribution 
between $Z/Z_{\odot} = 10^{-1}$ and $10^{-2}$ cases,
the temperature along the filamentary region becomes higher
in the lower metallicity case.
This strengthens the pressure support of the shock compressed region, 
which counteracts shock-compression and the contraction due to the self-gravity 
delaying the onset of star formation.
The turbulent motion is thus thermalized and quickly decays in the lower metallicity cases.
We can also see that the onset of star formation is delayed as the metallicity decreases.
The first protostar formation takes place at $t = 0.85t_\text{ff}$ when $Z/Z_{\odot}=10^{-1}$, 
while it takes place at $t = 3.9t_\text{ff}$ when $Z/Z_{\odot}=10^{-6}$.
The decay of the turbulent motion allows  
more mass to accumulate at the center 
and a massive gas disk develops.

Fig.~\ref{fig_vturb} shows the radial profile of the turbulent Mach number $v_\text{turb} / c_\text{s}$ at the time of formation of the first protostar for each case.
Here, the turbulent velocity is defined as 
the remainder after subtracting non-turbulent motion, i.e., the radial and circular motions, which are respectively caused by the gravitational collapse and initial cloud rotation, from the actual velocity field. 
To do this, we define the radial and circular motion at each mass shell at the given radius.
The radial and circular velocities are defined as 
the velocity toward the cloud center and that around the rotation axis, respectively. 
The orientation of the rotation axis is chosen to be that of the angular momentum vector at each mass shell.
In Fig.~\ref{fig_vturb} we average the turbulent velocity in the spherical shell at a given distance from the cloud center.
When $Z/Z_\odot \lesssim 10^{-3}$, the turbulent motion is sub- or trans-sonic inside $10^4~$au, 
while it becomes super-sonic at outer radii.
This causes a filamentary and axisymmetric structures in regions of low density as seen in Fig.~\ref{fig_snapshot}
at a few $10^4~$au from the cloud center.
When $Z/Z_\odot=10^{-2}$, 
the turbulence enters the super-sonic regime on scales $r\gtrsim 10^3~$au, much smaller than at lower metallicity. 
We can see this turbulent motion creates 
a filamentary structure in the central dense core in Fig.~\ref{fig_snapshot}.
When $Z/Z_\odot=10^{-1}$, 
the turbulent velocity becomes super-sonic already at $r\gtrsim 10~$au.
This means that the turbulent motion cascades toward much smaller scales than in the lower metallicity cases.

Fig.~\ref{fig_rhoT_hist} shows the density versus temperature diagrams for the different metallicities 
at the onset of star formation.
We can again see the general trend that the gas temperature decreases with metallicity.
The solid line shows the temperature evolution obtained from one-zone calculations,
where the cloud collapse is assumed to proceed at the rate of the free-fall time $t_\text{ff}$, 
i.e. $\dot{\rho} / \rho = 1/t_\text{ff}$, 
where $\rho$ is the density and $\dot{\rho}$ is the time derivative of the density.
Comparing the cloud temperature with those obtained by the one-zone models,
we can understand how turbulence changes the temperature evolution.
Turbulence affects the gas temperature in different ways depending on metallicity:
the gas temperature increases when $Z/Z_{\odot} \gtrsim 10^{-3}$, while it decreases when $Z/Z_{\odot} \lesssim 10^{-4}$.
In the higher metallicity cases,
shock heating results in higher temperature at $n \lesssim 10^{6}~\mathrm{cm}^{-3}$.
On the other hand, when $Z/Z_{\odot} \lesssim 10^{-4}$, the gas has lower temperature at $n\lesssim 10^{8}~\mathrm{cm^{-3}}$ since the turbulence delays the cloud collapse and gives the gas more time to cool, lowering the cloud temperature compared to that in the one-zone model.
The temperature evolution almost converges to those in the one-zone model at $n \gtrsim10^{8}~\mathrm{cm^{-3}}$
at any metallicity.

Note that when $Z/Z_\odot = 10^{-4}$ and $10^{-5}$,
the temperature is slightly higher 
than that expected from the one-zone model
in the density range
$n\sim 10^{12}$--$10^{14}~\mathrm{cm^{-3}}$,  
where the gas temperature is determined by the balance between the dust cooling and the adiabatic heating. 
In our simulations, the collapse indeed proceeds faster than the free-fall time 
assumed in the one-zone model (by a factor $1.5$--$2$ at $n\gtrsim 10^{10}~\mathrm{cm^{-3}}$),
leading to higher adiabatic heating rate and thus higher temperature than in the one-zone model.

Fig.~\ref{fig_cooling_rate} shows 
the time evolution of the cooling/heating rates at the cloud center 
as a function of the central density $n$ for different metallicities.
Shown are the cooling/heating rates averaged over the region within the Jeans length from the cloud center.
In the high metallicity cases of $Z/Z_{\odot}=10^{-1}$ and $10^{-2}$,
cooling is dominated by the C~{\sc ii} and O~{\sc i}  fine-structure line cooling
in the low-density regime where 
$n \lesssim 10^{6}~\mathrm{cm}^{-3}$, while, 
with metallicity $Z/Z_{\odot} \lesssim 10^{-3}$,
HD cooling dominates over the less efficient metal cooling.
The HD abundance, which is determined by the balance of the following two reactions,
\begin{align}
\mathrm{D}^{+} + \mathrm{H}_{2} \longrightarrow \mathrm{HD} + \mathrm{H}^{+}, \label{eq::HDformation}\\
\mathrm{HD} + \mathrm{H}^{+} \longrightarrow \mathrm{D}^{+} + \mathrm{H}_{2}, \label{eq::HDdestruction}
\end{align}
becomes higher at lower temperature. 
Although the H$_2$ abundance is lower for the lower metallicity gas due to the smaller contribution by the dust-catalyzed formation reaction, the temperature evolution at low densities, where H$_2$ is the dominant coolant, does not show a strong dependence on metallicity. 
This is because delayed collapse due to inefficient cooling at low metallicity lowers the adiabatic heating rate, 
allowing a longer time for H$_2$ formation, which enhances the cooling rate.
Once the temperature falls below $\la 150$K, most of D is locked up in HD and the gas cools down further to several tens of K by HD cooling.
This occurs for all the metallicities in our calculation unlike in the one-zone model in which the collapse rate does not depend on the cooling rate \citep[e.g.][]{Omukai+2005}.
We can in fact see that the temperature distribution is similar at $n \lesssim 10^{8}~\mathrm{cm}^{-3}$ for all the $Z/Z_{\odot} \lesssim 10^{-3}$ models.
This also explains why the gas temperature for $Z / Z_{\odot} \lesssim 10^{-3}$ becomes much smaller 
than that obtained by the one-zone calculation, where HD is not formed efficiently due to the assumption of the short free-fall collapse time-scale.
\citep[e.g.][]{Ripamonti2007, Hirano+2014, Chiaki+2020}.

Once the density becomes $n\gtrsim 10^{5}$--$10^{6}~\mathrm{cm}^{-3}$, the cooling rates by fine-structure and molecular lines become inefficient as the level population approaches the local thermodynamic equilibrium.
When $Z/Z_{\odot} = 10^{-1}$ and $10^{-2}$, temperature evolves almost isothermally with $\sim10~$K by dust thermal emission until it becomes optically thick at $n\gtrsim 10^{10}~\mathrm{cm^{-3}}$.
When $Z/Z_{\odot} \lesssim 10^{-3}$, the temperature increases up to a few $100~$K at $n\sim 10^{8}~\mathrm{cm}^{-3}$ due to the H$_{2}$ formation heating.
This is initially balanced by H$_{2}$ cooling
but soon after the cooling by dust thermal emission dominates, causing the temperature evolution minima when the central density is around $10^{10}$ -- $10^{14}~\mathrm{cm^{-3}}$.

Note that we have here neglected the formation of molecules such as CO, OH, and H$_{2}$O,
and the associated cooling, which becomes important 
when $Z/Z_{\odot} = 10^{-4}$--$10^{-3}$ \citep[e.g.][]{Omukai+2005, Chiaki+2016}.
Cooling by such molecular lines becomes important at $n \sim 10^{6}~\mathrm{cm^{-3}}$
and slightly modifies the temperature around this density.

\begin{figure*}
	\centering
		\includegraphics[width=18.cm]{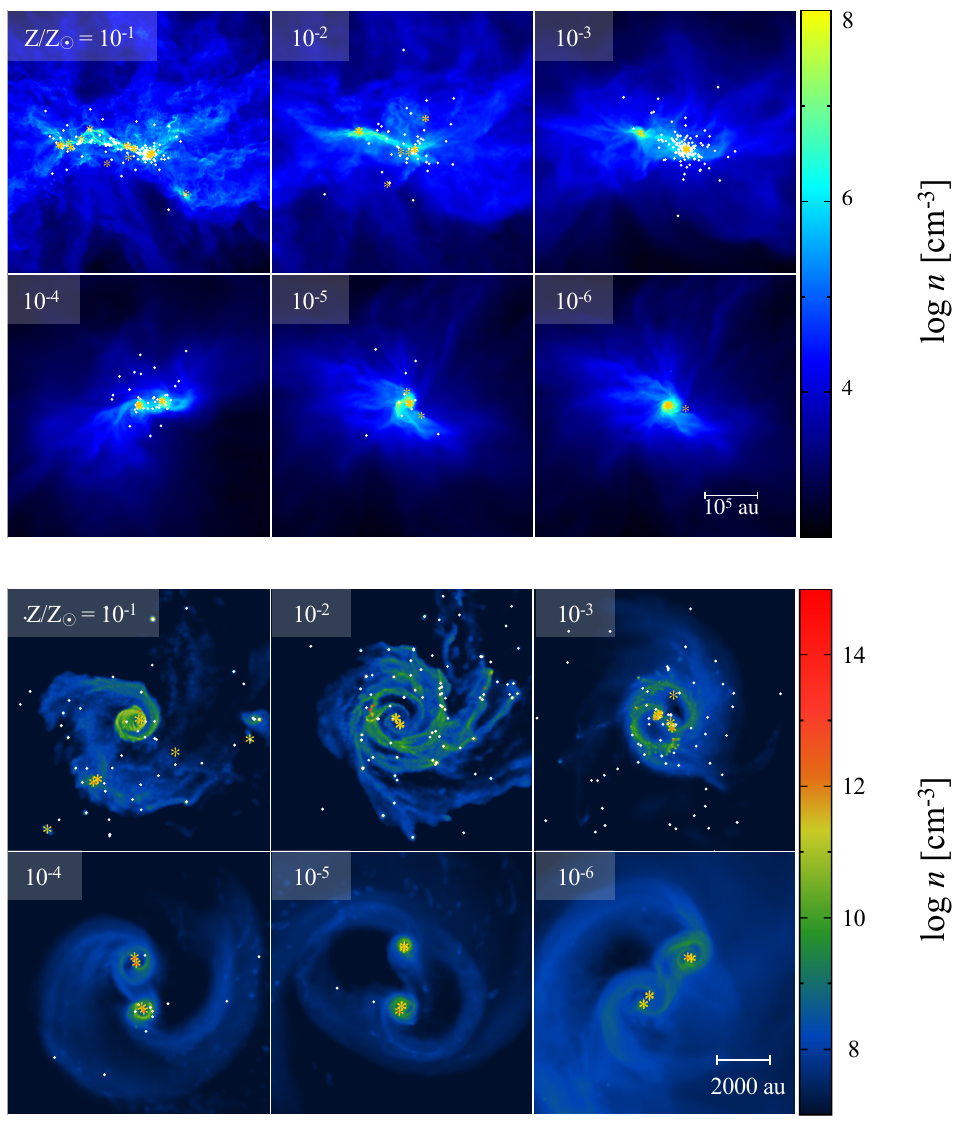}
		\caption{Top: the projected density distribution along z-axis when the total stellar mass reaches $150~M_{\odot}$ for different metallicity models.
		Yellow asterisks and white dots represent stars with masses larger than and smaller than $1~M_{\odot}$, respectively.
		Bottom: at the same epoch, the face-on view of the disks around the most massive stars or multiple stellar system.}
		\label{fig_snapshot_m150}
\end{figure*}

\begin{figure}
	\centering
		\includegraphics[width=8.cm]{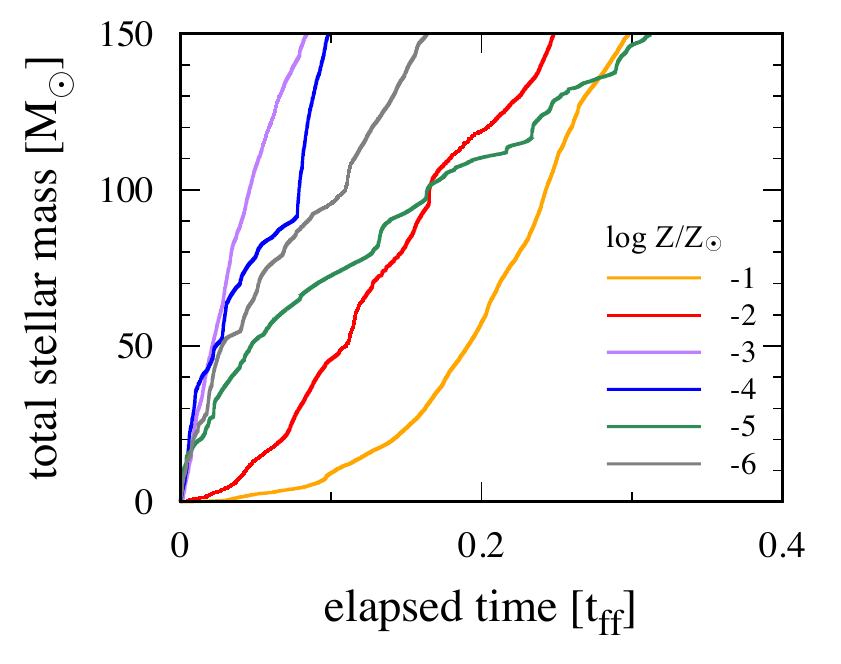}
		\caption{Time evolution of the total stellar mass $M_\text{tot}$
		for $\log Z/Z_{\odot} = -1$ (yellow), $-2$ (red), $-3$ (purple), $-4$ (blue), $-5$ (green), and $-6$ (grey).
		We set the time origin to be the time when the first protostar forms in the cloud.
		Time is normalized by the free-fall time $t_\text{ff}$ of the initial cloud core,
		where $t_\text{ff} = 4.7 \times 10^{5}$yr.
		}
		\label{fig_mtot_evolution}
\end{figure}

\begin{figure}
	\centering
		\includegraphics[width=8.cm]{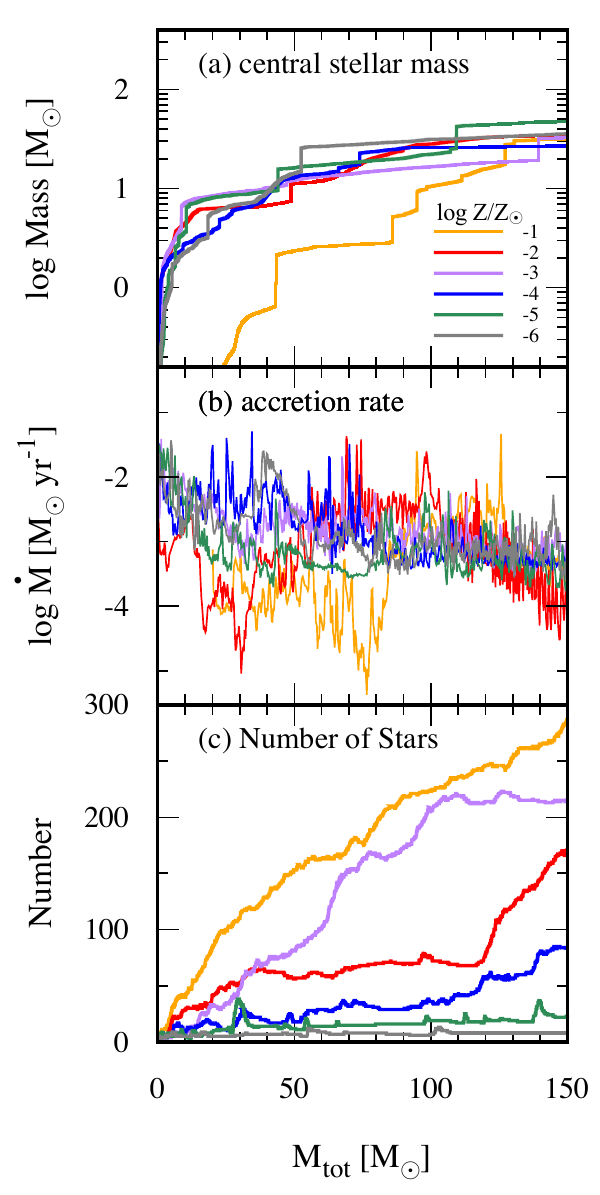}
		\caption{Evolution of 
		(a) the mass of the central stars,  (b) the mass accretion rate onto the central stars, and (c) the number of the stars formed in the cloud
		as a function of the total stellar mass $M_\text{tot}$
		for $\log Z/Z_{\odot} = -1$ (yellow), $-2$ (red), $-3$ (purple), $-4$ (blue), $-5$ (green), and $-6$ (grey).
		Since $M_\text{tot}$ increases with time (Fig.~\ref{fig_mtot_evolution}),
		the horizontal axis represents a time sequence from left to right.
		Here, we define the central star as the most massive star at $M_\text{tot}=150~M_{\odot}$.
		}
		\label{fig_mass_evolution}
\end{figure}

\begin{figure}
	\centering
		\includegraphics[width=7.5cm]{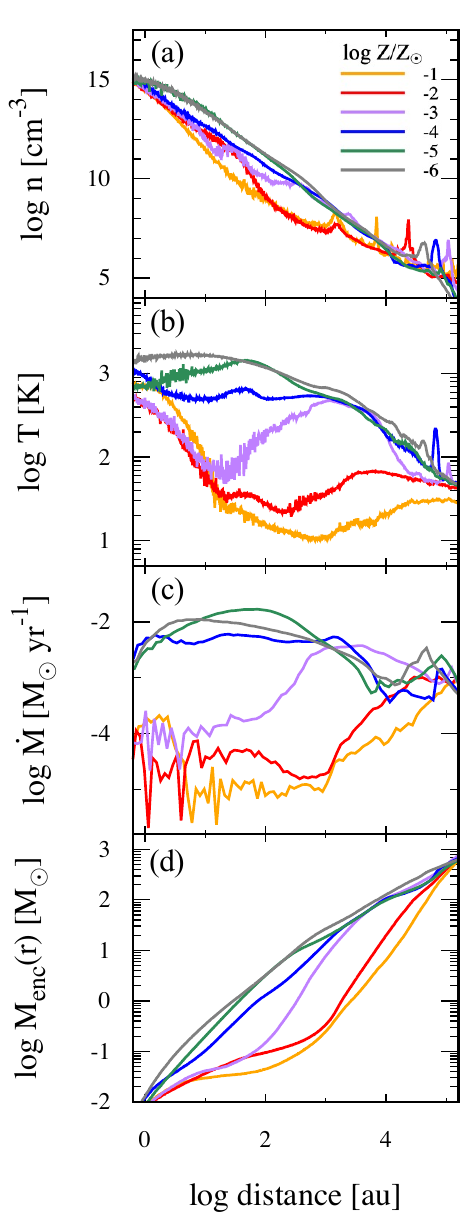}
		\caption{Radial profiles of (a) the density, (b) the temperature, (c) the inflow rate, 
		and (d) the enclosed mass inside the given radius
		at the moment when the star is formed.
		Different colors show the profiles with 
		$\log Z/Z_{\odot} = -1$ (yellow), $-2$ (red), $-3$ (purple), $-4$ (blue), $-5$ (green), and $-6$ (grey).}
		\label{fig_profiles}
\end{figure}

\subsection{Formation and evolution of the protostellar system}
After the onset of star formation, a number of protostars are formed in the course of our simulation.
The top panels of Fig.~\ref{fig_snapshot_m150} show the positions of the protostars for the different metallicities,
overplotted on the density distribution when in each simulation the total protostellar mass reaches $150~M_{\odot}$.
Protostars with masses larger (smaller) than $1~M_{\odot}$
are indicated with yellow asterisks (white dots, respectively).
When the metallicity is $0.1~Z_{\odot}$, the turbulent motion induces the formation of a filamentary structure, which fragments into protostars by gravitational instability.
At this metallicity, the filament vigorously fragments into small clumps,
allowing the formation of protostars throughout
the scale of the cloud core.
As the metallicity decreases, however, 
a filamentary structure can hardly develop
since the turbulent motion decays due to inefficient cooling.
This makes the formation sites of the protostars more concentrated.

The bottom panels of Fig.~\ref{fig_snapshot_m150} show the density distribution 
around the most massive protostars.
Here, we determine the orientation of the rotation axis around the most massive protostar
by calculating the angular momentum of the gas with $n>10^{8}~\mathrm{cm^{-3}}$
and show the face-on view of the gas distribution around the protostars.
We can see that a circumstellar disk forms at all values of the metallicity,
while the density structures are very different.
When the metallicity is smaller than $10^{-4}~Z_{\odot}$,
the gas disk surrounds the central massive binary 
or the multiple stellar system,
which is gravitationally stable at this epoch.
In these systems, gravitational instability recurrently operates
when the disk becomes massive owing to mass accretion from the envelope,
which induces fragmentation.
The fragment grows in mass via accreting the surrounding gas and
grows into massive protostars.
As a result, hierarchical binary systems form
as seen in simulations of star formation in primordial environments 
\citep{Chon+2018, Chon+2019, Susa2019, Sugimura+2020}.
The effect of dust thermal emission is only seen 
very close to the massive protostars within $\lesssim 100~$au,
which produces small fragments as shown in the $Z/Z_{\odot} = 10^{-4}$ case
since dust cooling becomes efficient only when $n \gtrsim 10^{11}~\mathrm{cm^{-3}}$.

In the models with $Z/Z_{\odot}=10^{-3}$ and $10^{-2}$,
the circumstellar disk becomes strongly unstable and 
widespread fragmentation leads to the formation of a large number of protostars.
This instability is mainly caused by dust cooling, which significantly reduces the gas temperature at $n \gtrsim 10^{8}~\mathrm{cm^{-3}}$.
Such dust-induced fragmentation leads to the formation of low-mass stars 
with typical masses of $0.01$--$0.1~M_{\odot}$,
reflecting the Jeans mass of the adiabatic cores \citep[e.g.][]{Omukai+2005,Chon&Omukai2020}.
Many of them show negligible increase in mass,
as they are ejected from the disk soon after their birth due to gravitational interaction with the central massive stars and with the spiral arms.
Once ejected, they wander in the low-density outskirt and the mass growth almost ceases in such cases.
When  $Z/Z_{\odot}=10^{-1}$, the circumstellar disk forms but its size is much smaller than those found at lower metallicities.
In this case, since strong turbulence survives until the later accretion stage, 
rotational motion is subdominant compared to the turbulent motion.
Due to the turbulent random motion, 
the angular momentum of the accreting gas does not align in the same orientation, 
and the disk structure is hard to form. 
The number of fragments formed in the disk is much smaller than in the lower metallicity cases.
Instead, fragmentation mainly takes place along the filament, rather than
in the circumstellar disk.

Fig.~\ref{fig_mtot_evolution} shows the time evolution of the total stellar masses
until they reach $150~M_{\odot}$.
We can see that the star formation time-scale is 
the shortest for the intermediate metallicity cases 
$Z/Z_{\odot} \sim 10^{-4}$ -- $10^{-3}$.
In such cases, rapid cooling in the high-density regime $n\sim 10^{10}~\mathrm{cm^{-3}}$
accelerates cloud collapse and induces vigorous star formation,
making the star formation time-scale very short.
In lower metallicity cases, inefficient cooling results in higher temperature and thus higher pressure support, 
which delays the cloud collapse and star formation.
On the other hand,
in higher metallicity cases,
the star formation time-scale increases again with increasing metallicity.
In this case, star formation is locally accelerated by the efficient cooling and compression by the turbulent motion.
This allows the first protostar to form earlier at higher metallicity:
the onset of star formation is $\sim 1t_\text{ff}$ for $Z/Z_{\odot}=0.01$--$0.1$,
while $\sim 1.7$ and $3t_\text{ff}$ for $Z/Z_{\odot}=10^{-3}$ and $10^{-4}$, respectively.
In the $Z/Z_{\odot}=0.01$--$0.1$ cases,
however, the turbulent motion increases the gas density only locally and then
only a small amount of gas accumulates around the cloud center at the time of the first protostar formation
(see also Fig.~\ref{fig_profiles}d).
This implies that a longer interval of time is required for accumulation of a gas reservoir
and subsequent episode of star formation in $0.1$--$0.01~Z_\odot$
with respect to the intermediate metallicity cases.

Fig.~\ref{fig_mass_evolution} (a) and (b) show
the evolution of the mass and the mass accretion rate for the most massive protostar at the end of our simulation, respectively.
The horizontal axis represents the total stellar mass in our simulated region
and can be interpreted as a time sequence going from left to right.
We can see that the mass growth histories of the central stars
are similar among the different metallicity cases
with the exception of the highest metallicity model.
The typical accretion rates at the end of the simulation converge to $\sim10^{-3}~M_{\odot}\mathrm{yr}^{-1}$, while the rates for $Z/Z_{\odot} \gtrsim 10^{-2}$ are initially smaller by an order of magnitude compared to the lower metallicity cases.
The similarity in the late time evolution comes from the fact that the most massive protostars are located at the center of the cloud.
The gas accumulates onto the cloud center, attracted by the sum of the gravity of the gas and stars in the system.
On the other hand, in the case of $Z/Z_{\odot}=0.1$, 
the sites of protostar formation is more controlled by persistent turbulent motions rather than the overall gravity of the cloud, 
and do not clustered around the cloud center.
This results in the difference in the later time accretion history.

Fig.~\ref{fig_mass_evolution}(c) shows the number of protostars
as a function of the total stellar mass.
There is a trend that the number of stars increases with increasing metallicity,
since efficient cooling induces the formation of a large number of fragments.
Between the models with $Z/Z_{\odot} = 10^{-3}$ and $10^{-2}$, however, this trend is reversed: a larger number of stars form when $Z/Z_{\odot} = 10^{-3}$ compared to the $Z/Z_{\odot} =10^{-2}$ case.
This is because the temperature at $n\sim 10^{8}~\mathrm{cm^{-3}}$ is much higher in $Z/Z_{\odot} = 10^{-3}$ than in the $10^{-2}$ model,
which makes the Jeans mass accordingly higher.

Fig.~\ref{fig_profiles} quantitatively explains why
the number of protostars is larger when $Z/Z_{\odot}=10^{-3}$ than in the $Z/Z_{\odot}=10^{-2}$ model,
showing the radial profiles
of the density (panel a), temperature (b), inflow rate (c), and enclosed mass inside the radius $r$ (d) as a function of the distance from the cloud center $r$ when the first protostar forms.
Here, we evaluate the radial profiles of the inflow rate
by $4\pi r^{2} \rho v_\text{inf}$ where $v_\text{inf}$ is the infall velocity toward the protostar,
averaged over the gas at the distance $r$.
The temperature profiles are quite different among the different metallicities.
At $r \lesssim 10^{4}~\mathrm{au}$, the temperature is much lower
in the high metallicity cases with $Z/Z_{\odot} \gtrsim 10^{-2}$ 
than in the low metallicity cases with $Z/Z_{\odot} \lesssim 10^{-4}$ due to efficient cooling at densities with $n \sim 10^{6}~\mathrm{cm}^{-3}$.
This makes the accretion rate, which is proportional to $T^{3/2}$ \citep{Larson1969, Shu1977}, much higher in the lower metallicity cases at this scales.
The case with metallicity $10^{-3}~Z_{\odot}$ is transitional in the sense that the temperature profile is close to that of the higher metallicity in the inner region with $r \lesssim 10^{2}~$au
but close to that of the lower metallicity in the outer region with $r \gtrsim 10^{3}~$au.
This causes a decrease in the inflow rate inward, from $10^{-3}~M_{\odot}\mathrm{yr}^{-1}$ in the outer region to $10^{-4}$ inside.
The gas accumulates at around $10^3$ au and the circumstellar disk becomes highly unstable,
leading to the formation of a larger number of protostars in this case
\citep{Tanaka&Omukai2014}.

Our results qualitatively agree with those obtained by \citet{Tanaka&Omukai2014},
where the circumstellar disk becomes gravitationally 
unstable when $Z/Z_\odot \sim 10^{-4}$--$10^{-3}$, 
although with the following quantitative difference: 
the disk is most unstable at $Z/Z_\odot = 10^{-4}$ in their analysis, 
while in our case it is more unstable at $Z/Z_\odot = 10^{-3}$ and $10^{-2}$.
This is caused by the different mass accretion rate from the cloud envelope onto the circumstellar disk.
The actual mass accretion rate in our simulation
is $10^{-3}$--$10^{-2}~M_\odot~\mathrm{yr^{-1}}$ in the metallicity range $Z/Z_\odot \lesssim 10^{-2}$ (Fig.~\ref{fig_profiles} c), higher than assumed by \citet{Tanaka&Omukai2014}, where 
$c_\text{s}^3/G \simeq 10^{-4}~M_\odot~\mathrm{yr^{-1}}$ at $Z/Z_\odot \gtrsim 10^{-3}$. 
This larger accretion rate leads to a more unstable circumstellar disk at $Z/Z_\odot = 10^{-2}$ and $10^{-3}$,
yielding a large number of low-mass stars in our study.

\begin{figure}
	\centering
		\includegraphics[width=8.8cm]{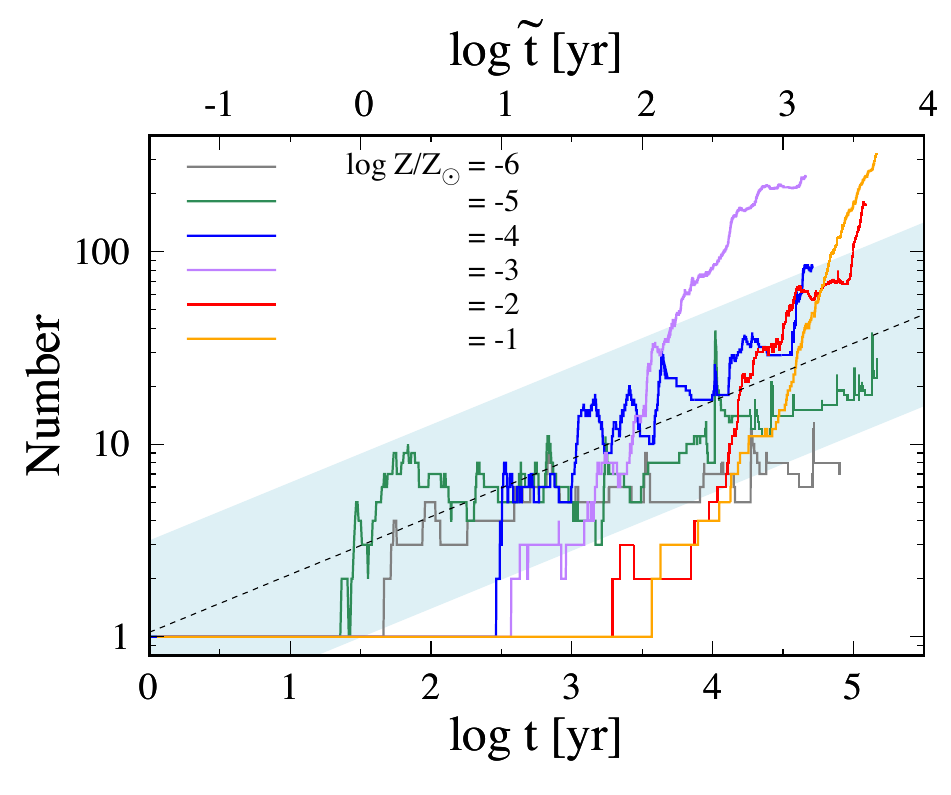}
		\caption{
		Time evolution of the stellar number for the metallicity with
		$\log Z/Z_{\odot} = -1$ (yellow), $-2$ (red), $-3$ (purple), $-4$ (blue), $-5$ (green), and $-6$ (grey).
		The black dashed line show the fitting formula (eq.~\ref{eq::stellarnumber}), proposed by \citet{Susa2019}.
		We overplot the blue shaded region, which is scatter of the stellar numbers 
		found in the studies on the primordial star formation.
		The upper horizontal axis shows the scaled time $\tilde{t}$, 
		which is defined in eq.~\eqref{eq::stellarnumber}.
		}
		\label{fig_Splot}
\end{figure}

Fig.~\ref{fig_Splot} presents the time evolution of the number of stars found in our calculation. 
Also shown by the dashed line is the relation in the case of the primordial star formation, 
which is proposed by 
\citet{Susa2019}, who compiled his own 
long-term simulation results and those by other authors: 
\begin{align} \label{eq::stellarnumber}
N(t) = 3 \left ( \frac{t_{\text{ff},\text{ad}}}{t_{\text{ff},\text{th}}} \frac{t}{1~\mathrm{yr}} \right )^{0.3}
\equiv 3 \left ( \frac{\tilde{t}}{1~\mathrm{yr}} \right )^{0.3},
\end{align}
where $t$ is the elapsed time since the first protostar is formed, 
$t_{\text{ff},\text{ad}}$ is the free-fall time at $n_\text{ad}=10^{19}~\mathrm{cm^{-3}}$, and
$t_{\text{ff},\text{th}}$ is the free-fall time at the density, where the sink particle is introduced.
When $Z/Z_\odot = 10^{-6}$ (grey), the number of stars is at the lower-edge of the shaded region.
When $Z/Z_\odot = 10^{-4}$ (green) and $10^{-5}$ (blue), the number roughly follows eq.~\eqref{eq::stellarnumber}.
In those cases, the stellar number sometimes decreases due to stellar mergers
while it increases by disk fragmentation afterward.
Although this causes fluctuations in the number of stars, its time average roughly obeys the relation of eq.~\eqref{eq::stellarnumber}.
Note that the decrease due to the mergers in this metallicity range is also reported by \citet{Shima&Hosokawa2021}, whose study follows the evolution in a few thousand years.
Our result suggests that the number of stars increases in a later stage than calculated in their study and behaves similarly to the primordial case.
Above $Z/Z_\odot = 10^{-3}$, the number of stars increases more steeply than in 
the lower metallicity cases and no longer obeys eq.~\eqref{eq::stellarnumber}.
In those cases, the circumstellar disk is highly unstable due to the dust cooling 
and a number of stars are ejected by the close stellar encounters (see Fig.~\ref{fig_snapshot_m150}).
In addition, fragmentation of the filament at larger scales ($10^4$--$10^5~$au) further boosts the number of stars in the case of $Z/Z_\odot \gtrsim 10^{-2}$.
Such effects as the stellar ejection and the fragmentation of the filament at the larger scale cause the deviation of the number of stars from eq.~\eqref{eq::stellarnumber} at $Z/Z_\odot \gtrsim 10^{-3}$.

\begin{figure*}
	\centering
		\includegraphics[width=17.cm]{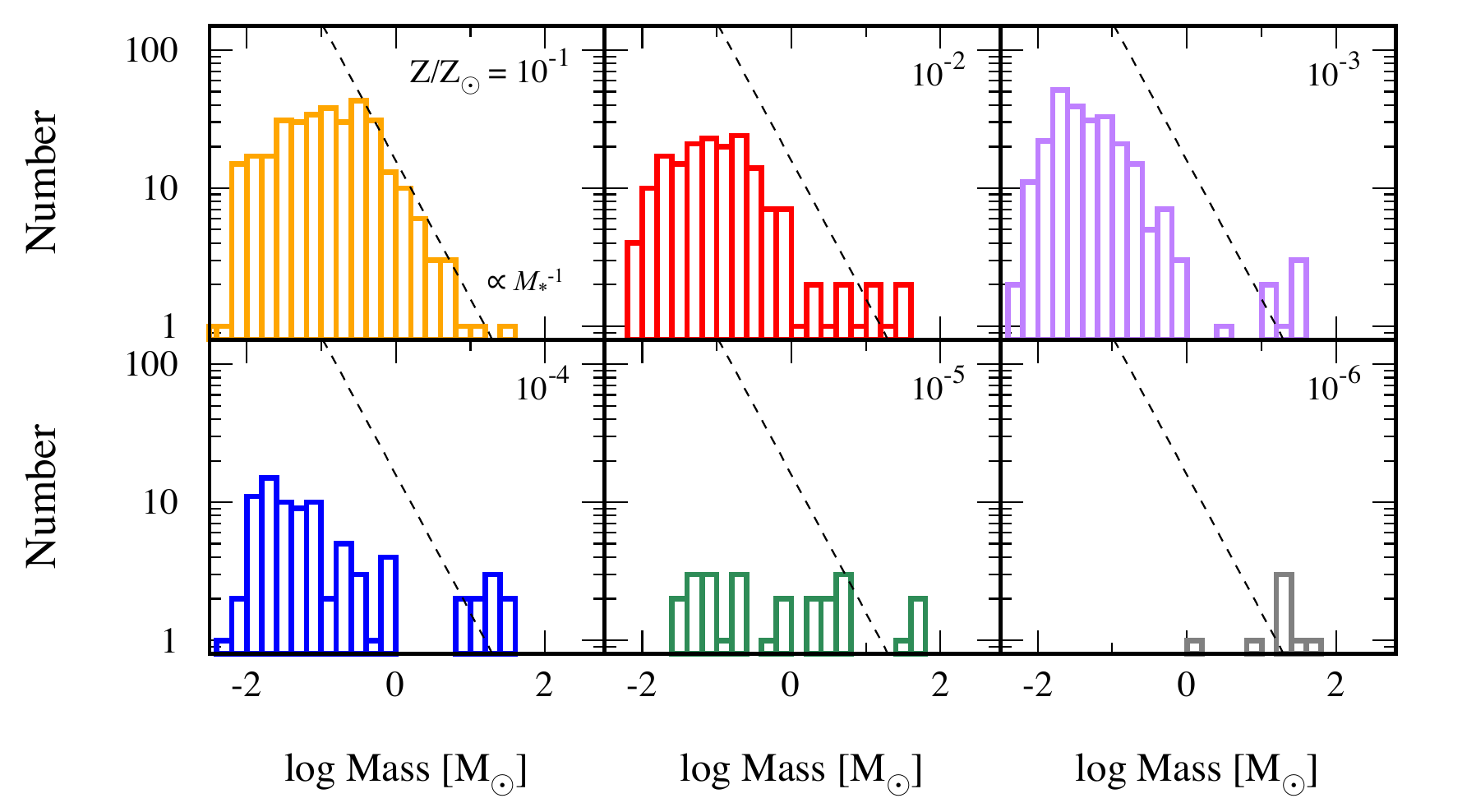}
		\caption{The mass distribution when the total stellar mass reaches $150~M_{\odot}$ for the different metallicity models. 
		The vertical axis represents the number of stars in each mass bin. 
		The black dashed lines show $\mathrm{d} N$/$\mathrm{d} \log M_{*}$ $\propto -1$,
		where $N$ is the number density in each mass bin, which has an equal width in logarithm.}
		\label{fig_mass_spectrum}
\end{figure*}

\begin{figure}
	\centering
		\includegraphics[width=8.5cm]{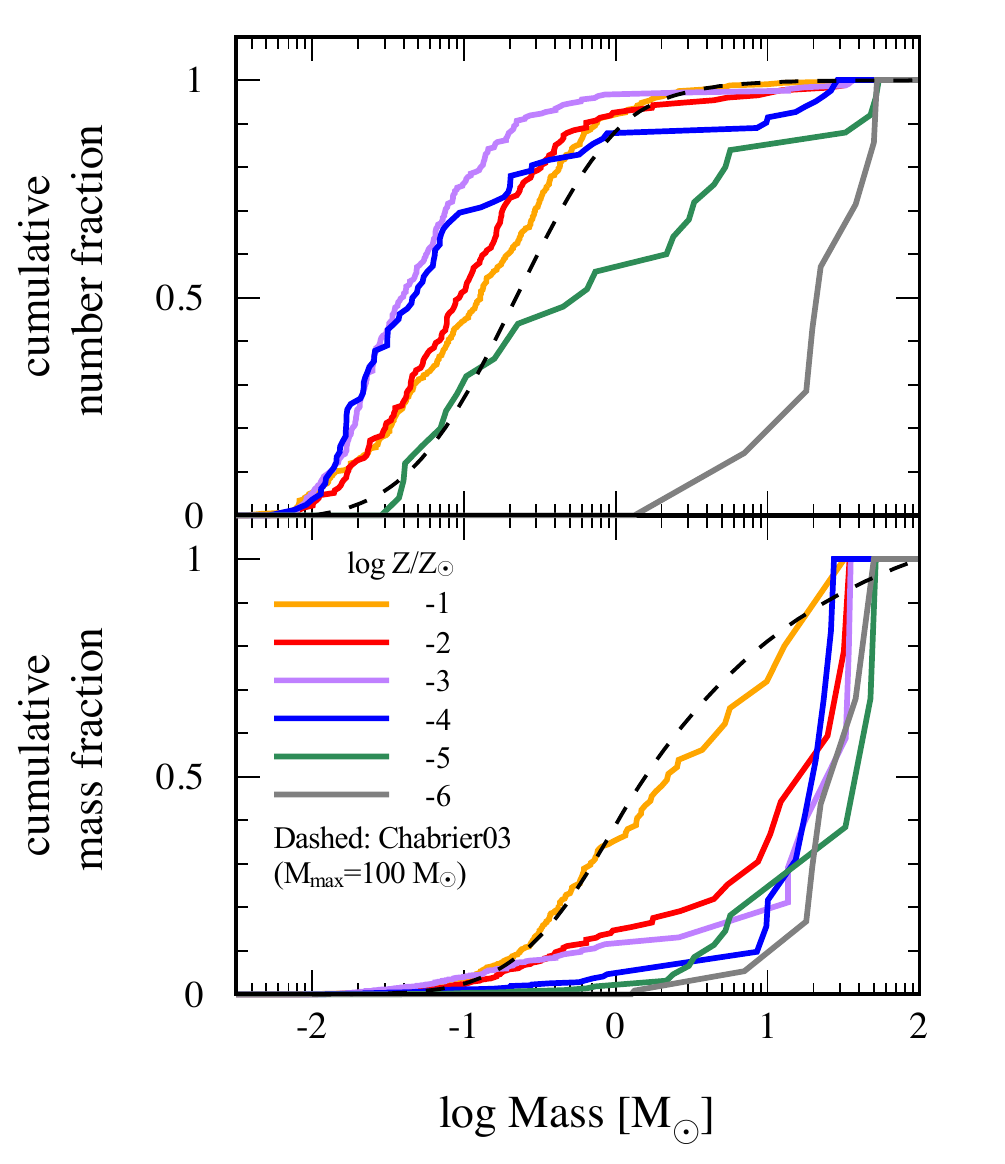}
		\caption{Cumulative number (top) and mass fraction (bottom) for
		$\log Z/Z_{\odot} = -1$ (yellow), $-2$ (red), $-3$ (purple), $-4$ (blue), $-5$ (green), and $-6$ (grey).
		Black dashed lines represent the cumulative distributions when we assume a Chabrier IMF \citep{Chabrier2003}
		with the maximum stellar mass of $100~M_{\odot}$.}
		\label{fig_cumulative_mass_spectrum}
\end{figure}

\begin{figure}
	\centering
		\includegraphics[width=8.5cm]{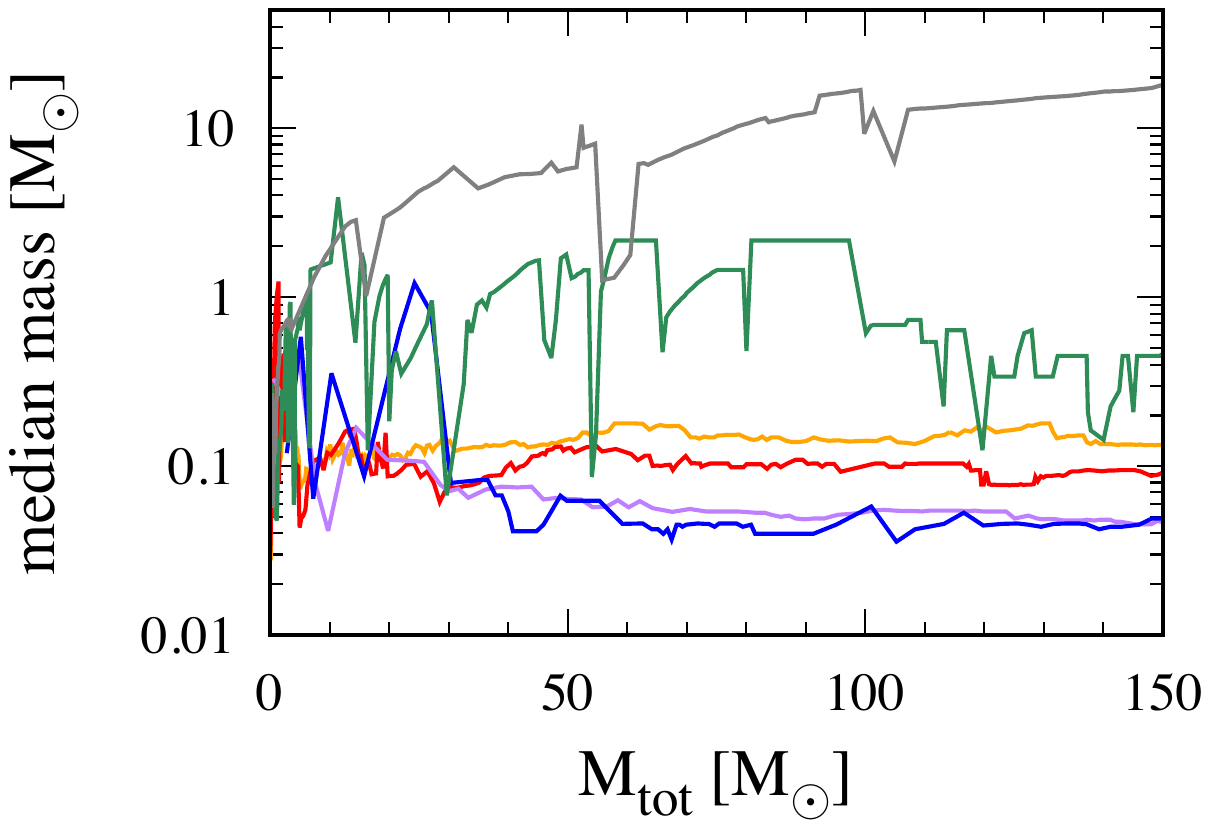}
		\caption{Evolution of the median mass as a function of the total stellar mass for
		$\log Z/Z_{\odot} = -1$ (yellow), $-2$ (red), $-3$ (purple), $-4$ (blue), $-5$ (green), and $-6$ (grey).
		}
		\label{fig_median_mass}
\end{figure}

Note that the thermal evolution in our study is different from that in \citet{Susa2019}, 
even for the case with $Z/Z_\odot = 10^{-6}$.
In our simulation, HD cooling becomes important and the temperature becomes smaller than $100~$K at
low density regions with $n \lesssim 10^8~\mathrm{cm^{-3}}$ (see Fig.~\ref{fig_rhoT_hist}).
This causes smaller gas infall rate toward the cloud center \citep[e.g.][]{Hirano+2015}
compared to the case where the HD cooling does not effective.
As a result, the circumstellar disk becomes smaller in mass and more stable against gravitational instability,
leading to a smaller number of stars.
In fact, the number evolution in $Z/Z_\odot = 10^{-6}$ (grey line) lies slightly outside the shaded region.

\begin{figure}
	\centering
		\includegraphics[width=8.5cm]{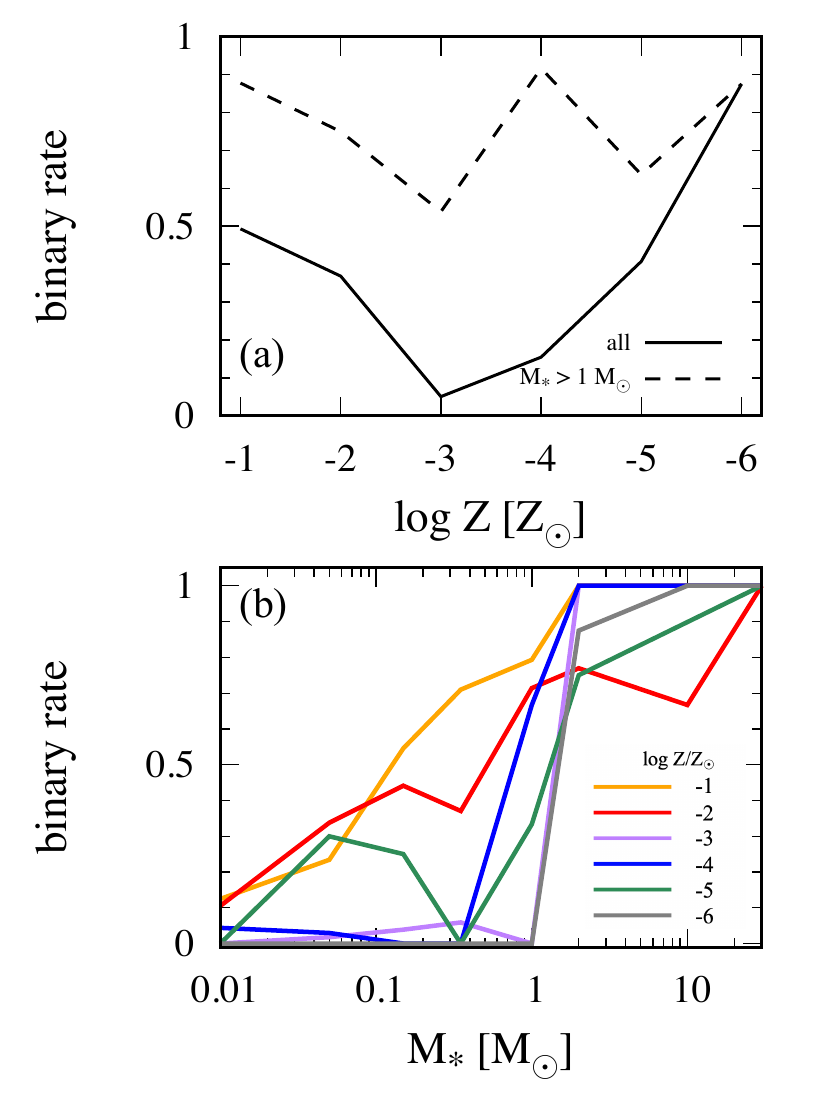}
		\caption{(a) Fraction that stars belong to a binary or multiple system
		for all the stars (solid) and massive stars with $M_{*} > 1~M_{\odot}$ (dashed)
		as a function of metallicity
		when the total stellar mass reaches $150~M_\odot$.
		(b) Fraction of stars belonging to a binary or multiple system as a function of metallicity
		for $\log Z/Z_{\odot} = -1$ (yellow), $-2$ (red), $-3$ (purple), $-4$ (blue), $-5$ (green), and $-6$ (grey).}
		\label{fig_binary_rate}
\end{figure}

\begin{figure*}
	\centering
		\includegraphics[width=15.cm]{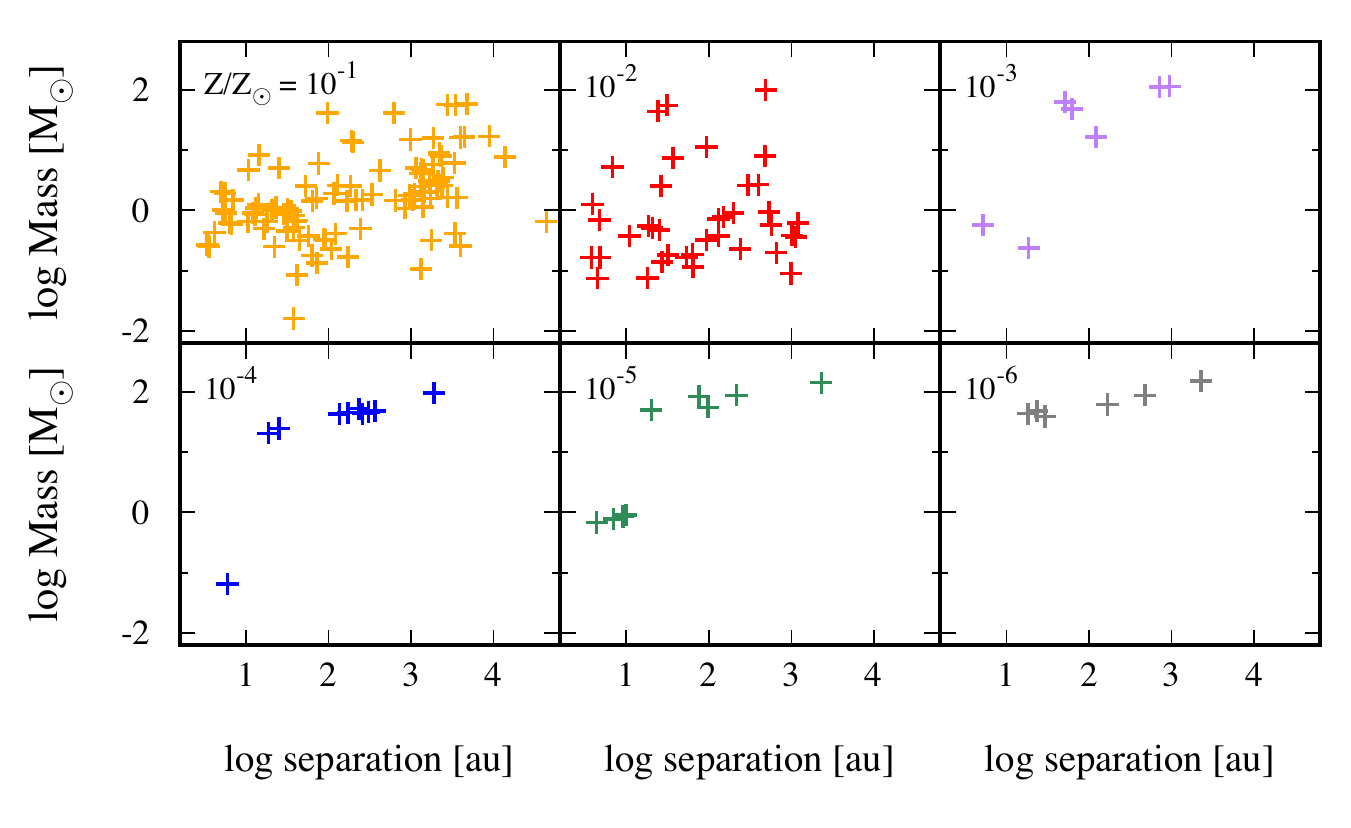}
		\caption{Total mass versus separation of binary systems for different metallicity models when the total stellar mass reaches $150~M_{\odot}$.}
		\label{fig_binary_mr_diagram}
\end{figure*}

\subsection{Mass function of the protostars}
Fig.~\ref{fig_mass_spectrum} shows the mass distribution of the protostars for the different metallicities when the total stellar mass reaches $150~M_{\odot}$. 
We can see that the distribution gradually shifts from top-heavy to Salpeter-like with increasing metallicity.
For example, when $Z/Z_{\odot}=10^{-6}$, all the protostars have masses larger than $1~M_{\odot}$ and the typical mass is several tens of solar mass.
When $Z/Z_{\odot}=10^{-5}$, the mass distribution becomes log-flat with the minimum stellar mass of $0.01~M_{\odot}$.
When $Z/Z_{\odot} \gtrsim 10^{-4}$,
a larger number of low-mass stars are formed due to dust-induced fragmentation.
This trend is consistent with the results of simulations by \citet{Dopcke+2011,Dopcke+2013},
where the number of low-mass stars increases with increasing metallicity.
The mass distribution for such low-mass stars is universal across a wide range of metallicity values: 
the number peaks at $0.01$--$0.1~M_{\odot}$ and declines in proportion to $M_{*}^{-1}$ at the massive end.
This universal distribution is consistent with those found in previous studies 
\citep{Bate2009, Safranek-Shrader+2014, Safranek-Shrader+2016, Bate2019}.
In addition to this universal profile, we can find 
a massive stellar component in the mass range of $1$--$50~M_{\odot}$ when $Z/Z_{\odot} \lesssim 10^{-2}$.
The number of stars associated with this component exceeds that expected from a simple extrapolation
from the lower mass end with the scaling $M_{*}^{-1}$.

To quantitatively compare the mass function for the different metallicities,
we plot in Fig.~\ref{fig_cumulative_mass_spectrum} 
the cumulative number (panel a) and mass distribution (b)
normalized by the total stellar number and mass, respectively.
The dashed line shows the cumulative fraction assuming the Chabrier IMF
with the maximum mass of $100~M_{\odot}$ \citep{Chabrier2003}.
The cumulative number fraction for $Z/Z_{\odot} \gtrsim 10^{-4}$ indicates that 
the number of stars is dominated by low-mass stars
with $M_{*} \lesssim 1~M_{\odot}$.
This is consistent with the results obtained by previous studies,
where the dust cooling induces the formation of a number of small stars
\citep{Clark+2008, Dopcke+2011, Dopcke+2013, Safranek-Shrader+2016}.
In terms of number fraction,
the critical metallicity is $Z_\text{crit} \sim 10^{-5}~Z_{\odot}$,
above which low-mass stars with $M_* < 1~M_\odot$
dominate the total number.

On the other hand, the cumulative mass fraction
indicates that the mass fraction of low-mass stars
is below 10\% in $Z/Z_{\odot} \lesssim 10^{-2}$ cases.
This value is much smaller than that expected from the Chabrier IMF (40\%),
indicating that the mass function is still top-heavy at these metallicities.
This means that the massive stars at the cloud center efficiently accretes mass,
while dust cooling produces a large number of small fragments.
When $Z/Z_{\odot} = 10^{-1}$,
the cumulative mass fraction becomes very close to the Chabrier IMF.
Therefore, in terms of the mass fraction,
the critical metallicity is $Z_\text{crit}/Z_{\odot} \sim 10^{-2}$ -- $10^{-1}$,
below which the mass function becomes top-heavy
compared to the present-day IMF.

Fig.~\ref{fig_median_mass} plots the time evolution of the median mass
as a function of the total stellar mass,
showing that it almost converges to $\sim0.1~M_{\odot}$
for $Z/Z_{\odot} \gtrsim 10^{-4}$.
Since the number of stars is dominated by the low-mass component,
this indicates that the stellar mass distribution of the low-mass stars 
does not evolve with time:
existing stars increase their mass by gas accretion,
while new low-mass stars continue to form, which makes the overall shape of the stellar mass distribution unchanged over time
\citep[e.g.][]{Bate+2003}.
This supports our interpretation of the mass distribution found by the simulations as the IMF.
Still radiative feedback from the stars can affect the gas distribution and mass growth rate,
as will be discussed in Sec~\ref{sec::feedbacks}.

\subsection{Statistical properties of stellar binaries and multiple systems}
In our simulation, some stars are formed as binary or multiple stellar systems with other stars. 
Statistical properties of the binaries are important in comparing our results with observations to verify our models.
Recent gravitational wave detection
have shown the existence of merging BHs with component mass larger than $30~M_{\odot}$
\citep[e.g.][]{GW150914,LIGO+O3a}.
Several authors insisted that such massive BHs are likely to be the end product of 
metal-poor or metal-free stars,
where reduced mass loss rate allows massive BHs to form
\citep[e.g.][]{Belczynski+2010, Kinugawa+2014, Schneider+2017, Marassi+2019, Spera+2019, Graziani+2020}.
Here, we discuss how the initial metallicity alters the stellar binary properties.

We define a pair of two stars to form a binary system if the total energy $E_\text{tot}$ is negative:
\begin{align}
E_\text{tot} = \frac{M_{1}v_\text{1}^{2}}{2} + \frac{M_{2}v_\text{2}^{2}}{2} -\frac{GM_{1}M_{2}}{r_\text{sep}}< 0,
\end{align}
where $M_{1}$ and $M_{2}$ are the masses of the pair of stars,
$v_{1}$ and $v_{2}$ are the velocities of the two stars relative to that of the center of mass,
and $r_\text{sep}$ is the separation of the stars.
In our simulation, stars often form a gravitationally bound group composed of more than two members,
so-called a higher-order multiple system.
Following \citet{Bate+2003}, we identify the hierarchically bound stellar groups 
according to the following procedure:
once we find a binary, we substitute the binary pair by a single stellar component,
whose position and velocity are those of the center of the mass of the binary.
We then find the bound pairs with this substituted component
and if it forms a binary with another star, 
we again substitute it with a single component.
By repeating the above process, we recurrently identify higher-order multiple systems.

In Fig.~\ref{fig_binary_rate}(a), the solid (dashed) line shows the probabilities 
that a star (a star with the mass larger than $1~M_{\odot}$, respectively) belongs to a binary system.
The binary rate decreases with decreasing metallicity when $Z/Z_{\odot} \gtrsim 10^{-3}$.
This trend can be interpreted as a consequence of the cloud morphology.
When $Z/Z_{\odot} = 10^{-3}$, the cloud collapses monolithically toward the cloud center,
and a number of stars are packed in the compact central region.
In such a situation, close stellar encounter frequently occurs, 
which ejects stars from the system or breaks the bound binary pairs,
reducing the binary rate in the lower-metallicity cases.
In contrast, when $Z/Z_{\odot} = 10^{-1}$, the filamentary structure develops and
stars form across a wide range in the cloud scales and close stellar encounter is
relatively rare compared to the lower-metallicity cases.
This environment is favorable for the formed binaries to avoid destruction by the stellar encounters, and 
the binary rate becomes higher than in the lower metallicity cases.

The binary rate of the stars with $M_{*} > 1~M_{\odot}$ 
seems to have little correlation with the cloud metallicity 
especially when $Z/Z_{\odot} \lesssim 10^{-4}$,
since it is mainly determined by the stochastic interaction 
between the massive stars at the cloud center.
Some massive stars are ejected from the cloud center via multi-body interaction,
which reduces the binary rate at $Z/Z_{\odot}=10^{-5}$.
Our result indicates that such massive stars residing at the cloud center 
usually form binary or multiple systems,
which is often observed in primordial star formation
\citep[e.g.][]{Stacy+2016,Chon+2018,Susa2019}.
On the other hand, the binary rate of all the stars strongly depends on the metallicity.
This is because the number of low mass stars becomes larger for higher metallicity
as dust cooling induces vigorous fragmentation.
Since low mass stars are easily ejected from the cloud and 
have smaller chance to form binary with other stars,
the binary rate becomes smaller at $Z/Z_{\odot}=10^{-3}$ 
than in lower metallicity environments.

Fig.~\ref{fig_binary_rate}(b) shows the binary rate as a function of the stellar mass.
Stars more massive than $2~M_{\odot}$ are mostly found in the binary system with its fraction somewhat depending on the metallicity.
Below $2~M_\odot$, the binary rate becomes small
due to the stellar ejection by the close stellar encounter, in which low-mass stars are more easily ejected than massive stars.
Once ejected, stars have little chance to make binaries with other stars,
which significantly reduces the binary rate of the ejected low-mass stars.
When $Z/Z_{\odot} \lesssim 10^{-3}$, this decrease is rather dramatic.
In this case, since formed stars are closely packed in the central compact region, 
the stellar encounter and the ejection happen more frequently.
As a result, most of the low-mass stars are ejected and the binary rate falls sharply at $M_* \lesssim 1~M_\odot$.
On the other hand, 
when $Z/Z_{\odot} \gtrsim 10^{-2}$,
the decrease of the binary rate is more gradual. 
In this case, formed stars are spatially
distributed over the entire cloud core
since stars are formed not only by the disk fragmentation but also by the filament fragmentation (see Fig.~\ref{fig_snapshot_m150}).
Owing to less frequent stellar encounter, low-mass stars formed in less dense region have more chance to make binaries before the ejection. 
Note that stars formed inside the disk, i.e., dense region, are ejected and hardly end up in binaries also in this case.

Fig.~\ref{fig_binary_mr_diagram} shows the total mass and separation of the binaries
for different metallicities.
When $Z/Z_{\odot} \lesssim 10^{-3}$,
it is hard to find a typical separation for the binaries.
In these cases, massive stars form hierarchical binaries 
(bottom panel of Fig.~\ref{fig_snapshot_m150}).
The separation of the closest binaries in those hierarchical systems is $\sim 10~$au.
Those tight binaries are embedded in hierarchical systems with
one or two companion(s) at separation of a few $100~$au,
forming triple or quadratic systems.
They also belong to higher-order multiple systems, with separation of a few $1000~$au.
Such hierarchical systems are often observed in simulations of primordial \citep{Susa2019}
or present-day star formation \citep[e.g.][]{Bate2009,Bate2019}.
A small number of low-mass binaries with mass of $0.1$ -- $1~M_{\odot}$ 
are also found, whose separation is smaller than $10~$au.
We find that those low-mass binaries have been ejected from the cloud center
due to the close encounter with massive stars,
which terminated the further mass growth.
Such ejected low-mass binaries tend to have small separations,
otherwise the binary is destroyed during the close encounter with massive stars.
When $Z/Z_{\odot} = 10^{-2}$, a massive hierarchical system resides at the cloud center 
surrounded by a large number of low-mass stars
(see the bottom panel of \ref{fig_snapshot_m150}).
This system is composed of two tight binaries with separation of a few $10~$au, forming a massive quadruple system as in the cases with lower metallicity.
Low-mass binaries have separation of a few -- $10^{3}~$au, which is not observed in the lower metallicity cases.
When $Z/Z_{\odot} = 10^{-1}$, 
a large number of low-mass tight binaries appear while massive binaries typically have larger separations.

We have also seen that tight binary systems form in hierarchical multiple systems.
Previous studies have shown that the separation of an isolated binary system increases as it acquires the angular momentum from the accreted matter
\citep{Susa2019, Chon+2019}.
In contrast, if a binary is embedded in a hierarchical system, 
the outer-most object preferentially acquires the angular momentum and increases its separation with the inner objects, while the separation of more inner systems does not grow over the simulated period as long as they do not experience a close encounter with another star.
This fact is consistent with previous studies on the primordial star formation,
in which the authors observed the formation of massive binary systems
\citep[e.g.][]{Stacy+2016, Sugimura+2020}.

Our results show that the formation of the massive tight binaries is expected
for all the metallicity range considered here.
Their separations are typically $\sim10~$au and
show little evolution with time. 
This supports the idea that merging BH binaries are 
formed in low-metallicity environments, including the primordial case.
Still the separation is too large for the binary to coalesce within the Hubble time 
via the GW emission, if it evolves into a BH binary
\citep{Peters1964}.
Since the minimum binary separation set by the 
numerical resolution is $\sim$au,
binaries with sub-au separation cannot be resolved in our calculation.
Since higher-order multiple systems are observed down to $10~$au scales, multiple systems with even smaller separation could be found in calculations
with higher spatial resolution of sub-au scale.
To investigate the possibility whether 
low-metallicity clouds yield merging binary BHs,
we need to further resolve the stellar radii and follow binary formation at sub-au separations.

\section{Discussion} \label{sec::discussion}
\subsection{Transition from top-heavy to present-day IMF}
We have studied the impact of metallicity on the stellar mass spectrum
and found that the number of low-mass stars ($M_* < 1~M_\odot$) increases with metallicity.
Our results are not only consistent with previous studies 
on the mass functions found in low-metallicity environments
\citep{Clark+2008, Dopcke+2011, Dopcke+2013, Safranek-Shrader+2016},
but also exemplifies a natural transition to the present-day Salpeter-like IMF
found by numerical studies of stellar cluster formation in the literature 
\citep{Bate+2003, Bonnell+2006, Bate2009, Krumholz+2012, Bate2019}.
Here, we briefly compare our findings with previous results and
discuss how the metallicity effect modifies the stellar mass spectrum.

Previous numerical studies have shown that 
in low-metallicity environments with $Z/Z_{\odot} \lesssim 10^{-4}$
the existence of a trace amount of metals leads to the formation of low-mass stars 
owing to fragmentation induced by dust cooling 
\citep{Clark+2008, Dopcke+2011, Dopcke+2013, Safranek-Shrader+2016}. 
For example, \citet{Dopcke+2013} have found that 
the stellar mass distribution is log-flat for $Z/Z_{\odot} \lesssim 10^{-5}$,
while the Salpeter-like power law distribution with the peak mass of $0.1~M_{\odot}$
develops when $Z/Z_{\odot} = 10^{-4}$.
Still, their calculations are extended only up to several hundred years and 
most of massive protostars are still vigorously accreting the mass at the end of the simulation,
so the mass distribution can still change with time.
Our results are consistent with their findings for this metallicity range,
if we focus on the mass distribution at the low-mass end, i.e., 
log-flat for $Z/Z_{\odot} \lesssim 10^{-5}$ while Salpeter-like for $Z/Z_{\odot} \gtrsim 10^{-4}$.

One big difference between our and previous results can be observed at the high-mass end of the mass distribution, i.e., the existence of massive stars with several $10~M_{\odot}$.
This massive component cannot be accounted for by a simple extrapolation from the Salpeter-like power-law tail developed at the lower masses.
This is a natural consequence of the fact that the stellar mass at the high-mass end is mainly determined by the cluster's gravitational potential.
The massive protostars preferentially grow in mass, since they reside at the bottom of the gravitational potential well,
as typically quoted in the ``competitive accretion'' model \citep{Bonnell+2006}.
Their masses are almost independent of the gas metallicity, which only alters the thermal state of the cloud and thus the typical mass of the fragments \citep[e.g.][]{Larson2005}.

When the metallicity is $Z/Z_{\odot}=0.1$, a highly filamentary structure develops,
which is often seen in the present-day stellar cluster formation in a turbulent cloud core
\citep{Li+2003,Bate+2003, Bonnell+2004, Jappsen+2005, Smith+2011, Krumholz+2012, Guszejnov+2018}.
This is also consistent with the morphology of the observed star forming regions by Herschel space observatory 
\citep[e.g.][]{Andre+2010, Konyves+2015, Marsh+2016}.
We have shown that dust cooling at $n \gtrsim 10^{6}~\mathrm{cm^{-3}}$ lowers the effective ratio of specific heat $\gamma$ below unity and causes the growth of the filamentary mode of the gravitational instability 
\citep[e.g.][]{Bastien1983,Inutsuka&Miyama1992, Hanawa&Matsumoto2000, Tsuribe&Omukai2006}.
Our results suggest that such a filamentary structure does not develop for $Z/Z_{\odot} \lesssim 10^{-3}$, since the cooling rate in shock compressed gas is small and boosted pressure support prohibits the growth of the filamentary mode.
This value of the critical metallicity is consistent with several observational evidences.
For example, the fraction of carbon-enhanced stars increases with decreasing metallicities, which can be explained by the IMF transition at $Z_\text{crit}/Z_\odot \sim 10^{-2.5}$
\citep{Suda+2013,Lee+2014}.
To reproduce the observed profiles of globular clusters, \citet{Marks+2012} suggests that
the top-heavy IMF is realized in low-metallicity environment with $Z/Z_\odot \sim 10^{-2}$ to $10^{-1}$.
They explain the trend that the lower-metallicity clusters have lower concentration of low-mass stars 
by considering the top-heavy IMF, which quickly depletes the cluster gas.

We claim here that monolithic collapse of the central massive core 
leads to formation of central massive stars,
and thus to a top-heavy IMF.
Once the turbulent motion decays, the filamentary structure no longer develops
as seen in the $Z/Z_{\odot} \lesssim 10^{-3}$ case.
In this case, the gas accumulates at the cloud center, forming a massive circumstellar disk.
Since a large amount of mass is confined to the small central region, massive stars in this region can efficiently accrete mass.
Dust cooling induces vigorous fragmentation inside the disk and produces a number of low-mass stars, while they are easily ejected from the system as they experience many-body interactions with other stars
\citep{Clark+2008, Dopcke+2013, Safranek-Shrader+2016, Chon&Omukai2020}
or migrate inward due to the interaction with the disk gas and merge with the central stars 
\citep[e.g.][]{Tanaka+2002, Chon+2019}.
This makes the massive stars more massive,
producing a high-mass stellar component distinctive from the distribution of the low-mass stellar component.

Survival of the turbulent motion is the key to form a filamentary structure \citep[e.g.][]{Krumholz+2012}.
In our study, we impulsively inject a turbulent energy at the initial collapse stage, which decays over the time-scale of the sound crossing time or the free-fall time of the cloud \citep[e.g.][]{Stone+1998,MacLow1999,Bate+2003}.
The turbulence can be continuously driven by the external forcing, such as the expansion of the H~{\sc ii}  region or injection of mechanical energy by SN explosions outside our simulation box.
How such forcing works in low-metallicity environments is still unclear.
Nonetheless, we expect inefficient cooling in the low-metallicity gas to strongly resist turbulent compression and to suppress the formation of filamentary structure even when the turbulent energy is continuously injected.
Such continuous forcing in low-metallicity environments would delay the cloud collapse 
and can lead to formation of more massive stars
compared to non-turbulent clouds
\citep[e.g.][]{McKee&Tan2003}.

Our results can also have strong impact on the chemical evolution of galaxies.
\citet{Larson1998} discussed the possibility that a top-heavy IMF in the early universe 
causes rapid chemical enrichment and thereby
reduces the number of low-metallicity stars.
This can be a clue to the so-called classical 'G-dwarf problem' \citep{Tinsley1980}, i.e., 
the number of low-metallicity stars are smaller than theoretical expectation 
in the closed-box model without gas inflow or outflow \citep{Salvadori+2007}.
\citet{Larson1998} also suggested that a top-heavy IMF in the early universe can explain a large amount of heavy elements observed in hot gas of galaxy clusters.
Effect of a metallicity-dependent IMF on the chemical evolution of galaxies has been 
studied by some authors.
\citet{Bennassuti+2017} showed that a top-heavy IMF in primordial minihalos successfully reproduces the metallicity distribution of the low-metallicity stars in the Galactic halo.
Other studies on Galactic chemical evolution showed that
a top-heavy IMF in low-metallicity environments solves the G-dwarf problem but fails to reproduce the observed abundance patterns \citep{Chiappini+2000, Martinelli+2000, Ballero+2006}.
The IMFs assumed in their studies, i.e., 
single power-law simple IMFs, however, 
are different from our finding, 
i.e., Salpeter-like IMFs plus massive components with several $10 M_{\odot}$ in low-metallicity environments.
Effects of the IMF derived in this paper
on galaxy formation and evolution is worth examining quantitatively in future studies.

\subsection{The impact of radiative feedback and magnetic field on IMF} \label{sec::feedbacks}
Protostars are still accreting the gas at the end of our simulation since we do not include any feedback effects from the forming stars, which can halt further mass accretion.
The ionizing radiation emitted from massive stars is one of the leading processes that disperse the surrounding gas and terminates mass accretion
\citep{Peters+2010, Dale+2012, Walch+2012, Geen+2018, He+2019, Fukushima+2020b}.
\citet{He+2019} conducted radiation hydrodynamics simulations and found that while the ionizing radiation decelerates the protostellar mass growth, the functional form of the mass distribution, such as typical stellar mass, slope of the high mass tail, shows little time evolution during their simulation.
They provided the following analytical fit to the total star formation efficiency $f_{*}$,
\begin{align}
f_{*} = 0.02 \left ( \frac{M_\text{cloud}}{10^{4}~M_{\odot}} \right )^{0.4} \left ( 1 + \frac{\bar{n}}{10^{3}~\mathrm{cm^{-3}}} \right )^{0.91},
\end{align}
where $M_\text{cloud}$ and $\bar{n}$ is the mass and mean density of the initial cloud core, respectively.
Using our cloud parameters of $M_\text{cloud}=1950~M_{\odot}$ and $\bar{n}=10^{4}~\mathrm{cm^{-3}}$, 
we obtain the total stellar mass $\sim170~M_{\odot}$, which is close to the stellar mass at the end of our simulation.

The impact of the ionization feedback becomes stronger with decreasing metallicity,
since the less efficient cooling inside the H~{\sc ii}  region increases the pressure in the ionized gas,
which easily evacuates the cloud gas \citep{He+2019, Fukushima+2020b}.
Our simulation also suggests that difference in the cloud metallicity changes the collapsing cloud morphology, 
which can result in difference in the feedback efficiency.
For example, when $Z \gtrsim 0.01~Z_{\odot}$, 
massive stars are spatially distributed over the entire cloud core.
Meanwhile, when $Z \lesssim 10^{-3}~Z_{\odot}$, massive stars are closely packed around the cloud center, which is surrounded by a gas disk (see Fig.~\ref{fig_snapshot_m150}).
High density in the disk can efficiently attenuate UV radiation from the central stars in the disk plane.
Since a large amount of mass is still supplied through the disk plane, the stars would continue to grow in mass
despite the UV radiation feedback
\citep[e.g.][]{McKeeTan2008,Hosokawa+2012}.
\citet{Fukushima+2020a} followed the formation of stars starting from a massive collapsing cloud
of $10^3~M_\odot$.
They found that the total mass of stars reach several hundred solar mass when $Z \lesssim 10^{-3}~Z_{\odot}$, meaning that the star formation efficiency is about $50$\%.
To further investigate the impact of metallicity on the star formation efficiency
in low-metallicity environments,
we should follow the propagation of UV radiation 
resolving down to the circumstellar disks around massive stars.

Heating due to non-ionizing photons is another important process called ``thermal feedback'', 
which can alter the protostellar mass distribution.
For stars with $M_{*} \lesssim 10~M_{\odot}$, 
the stellar luminosity is dominated by the accretion luminosity \citep{Hosokawa+2009, Fukushima+2018}.
This heats up the dust grains and finally increases the gas temperature, prohibiting further fragmentation and thus formation of low-mass stars 
\citep{Bate2009, Omukai+2010, Myers+2011, Bate2012, Krumholz+2012, Bate2019}.
Since thermal feedback does not prevent converging accretion flows from reaching the central massive stars, 
its main effect is to reduce the number of low-mass stars,
increasing the typical protostellar mass.
In our simulation, the fractional number of brown dwarfs are larger than that predicted by the Chabrier IMF (Fig.~\ref{fig_cumulative_mass_spectrum}a).
This over-abundance of brown dwarfs can be potentially mitigated by the inclusion of thermal feedback \citep{Bate2009, Krumholz+2012}.

Magnetic fields can affect the stellar mass distribution by launching outflows and then reducing the mass accretion rate 
\citep{Machida+2008b, Federrath+2014, Kuruwita+2017, Matsushita+2018}.
The magnetic force can also prevent gas fragmentation, thereby reducing the number of low-mass stars
\citep{Mellon&Li2008, Machida+2011b, Myers+2013}.
Still, the magnetic-field strength in low-metallicity environments is very unclear.
Several magnetic-field generation mechanisms have been advocated,
such as the small-scale dynamo \citep[e.g.][]{Schleicher+2010,Schober+2012}
or the amplification during the expanding SNe and H~{\sc ii}  shells \citep[e.g.][]{Koh+2016},
some of which predict that magnetic fields can be created even in primordial environments.
Recent simulations follow the formation of a star cluster starting from a magnetized cloud in the present-day environments \citep{Myers+2014, Cunningham+2018}.
They conclude that the impact of magnetic fields on the IMF is rather small compared to radiative feedback,
leading to a small increase in the peak mass by less than a factor of three.
Assuming that the magnetic fields are weaker in low-metallicity environments at high redshift
due to the limited time for magnetic field amplification,
we may expect the fields to play only a subdominant role in shaping the IMF.
On the other hand,
the field may modify the mass distribution at the high-mass end by making protostellar disk fragmentation inefficient by removing 
angular momentum and allowing formation of more massive stars at the cloud center
\citep[e.g.][]{Machida+2008b,Kuruwita+2017,Sadanari+2021}.
If so, this would have significant impact on the efficiency of stellar feedback, for example, 
increasing the frequency of pair instability supernovae.

\subsection{Initial conditions for low-metallicity star formation}
In this paper, we have initiated our calculation from the same gas profile regardless of the metallicity, i.e., the Bonnor-Ebert sphere with the central density $10^4~\mathrm{cm^{-3}}$ and temperature $200~$K.
With high enough metals, i.e., $\ga 10^{-2}~Z_\odot$,
the temperature quickly  drops by metal-line cooling after the onset of the calculation. 
As a result, the temperature profiles become very different depending on the initial metallicity. 
Once the thermal energy is dissipated away, the thermal evolution plays only a subdominant role in shaping the overall structure of the cloud.
Instead, the turbulent motion determines the cloud structure and dynamics.
Our results demonstrate the importance of turbulence in low-metallicity environments:
thermal evolution mostly controls the dynamics at $Z/Z_\odot \lesssim 10^{-3}$, while
turbulent motion is more important at $Z/Z_\odot \gtrsim 10^{-2}$.

In the lowest-metallicity case with $Z/Z_\odot = 10^{-6}$, 
a realistic initial condition would resemble what is found by cosmological simulations
for first star formation in a minihalo. 
In this study, we have adopted a turbulent cloud as the initial condition for the calculation, where the turbulent energy is slightly ($\lesssim 2$) larger than in typical minihaloes \citep[e.g.][]{Greif+2012}.
This difference results in smaller number of fragments in our calculation than in previous studies (cf. Fig.~\ref{fig_Splot}).
Here, the delayed collapse due to turbulent motion allows the gas to cool more by HD, which leads to smaller accretion rate onto the central protostars after their formation, as the accretion rate depends on the prestellar temperature as $\propto T^{3/2}$. 
Consequently, the circumstellar disks around the protostars are less massive and fragment less frequently, producing a smaller number of stars when turbulent initial conditions are adopted.

Still, appropriate initial conditions for  low-metallicity star formation are currently poorly understood.
We have little knowledge about the properties of turbulent motions, gas density structures, etc.
In the local Universe, star formation is known to take place 
inside turbulent molecular clouds with density $100$ -- $10^4~\mathrm{cm^{-3}}$.
Formation process of molecular clouds has been  studied in such scenarios as gravitational instability of dense gas in galactic disks \citep[e.g.][]{Hopkins2012, Dobbs&Baba2014, Kruijssen2014} or
thermal instability of colliding flows of warm diffuse atomic medium
\citep[e.g.][]{Koyama&Inutsuka2002, Hennebelle+2008, Inoue&Inutsuka2009, Inutsuka+2015}.
With this in mind,
as the initial condition of our simulation
we took a turbulent cloud with turbulent energy comparable to gravitational energy motivated by observations of local molecular clouds
\citep{Larson1981, Heyer+2004}.
Applicability of such initial condition also to low-metallicity environments needs further justification by 
both future observations and theoretical inspection.
Recently, \citet{Kalari+2020} observed the star-forming region Magellanic Bridge C with metallicity $\sim 1/5~Z_\odot$ by ALMA and identified an associated filamentary structure.
Such morphological study on star forming regions 
will give us some insight on the initial conditions for star formation in low-metallicity environments.
Theoretical studies are also required to reveal the formation process of star-forming clouds. 
For example, clump formation by thermal instability in low-metallicity environments has been studied by \citet{InoueOmukai2015}.

We have followed the stellar mass evolution until the total stellar mass reaches $\sim 150~M_\odot$,
at which epoch the radiation feedback will operate (see section~\ref{sec::feedbacks}).
Although our simulation is extended longer than the previous studies 
\citep{Clark+2008,Safranek-Shrader+2016, Chiaki+2020, Shima&Hosokawa2021},
the total stellar mass is still small compared to observed star clusters, 
whose masses range from $\sim 100$ to $10^5~M_\odot$ 
\citep[e.g.][]{Krumholz+2019}.
A more massive stellar cluster would be expected to form, if we start from initially a more massive cloud.
\citet{Marks+2012} suggested that the IMF can be more top-heavy for increasing initial cloud mass or initial density.
This explains why massive globular clusters tend to have lower concentration \citep{DeMarchi+2007}, 
because a top-heavy IMF dissipates the cluster gas and makes the cluster less concentrated.
Our results suggest that even in more massive systems,
inefficient cooling dissipates turbulent motion
and leads to a top-heavy mass distribution.
The critical metallicity which marks the transition from the top-heavy to Salpeter-like IMF
can also depend on the initial cloud mass.

\subsection{Other important effects}
There are several important effects we have not considered in our simulation.
One is the warmer CMB in the high-$z$ universe,
which sets the minimum dust and gas temperatures to be $2.72(1+z)~$K.
This makes the temperature evolution almost isothermal at the CMB temperature when the gas density is $n \lesssim 10^{10}~\mathrm{cm^{-3}}$, and the metallicity has smaller impact on the cloud evolution because the gas cannot cool below the CMB temperature \citep{Smith+2007, Jappsen+2009, Meece+2014, Bovino+2014, Safranek-Shrader+2014}.
\citet{Riaz+2020} investigated how the CMB temperature affects gas fragmentation
and thus the stellar mass distribution.
They found that higher CMB temperature suppresses fragmentation, resulting in a top-heavy mass distribution.
Still the temperature evolution should not be completely isothermal, since the formation heating of hydrogen molecules increases 
the gas temperature to a few $100~$K \citep{Schneider&Omukai2010}.
We will pursue this point in a future paper.

In our simulation, we impose several assumptions about the properties of dust grains,
i.e. the dust-size distribution \citep{Mathis+1977},
the composition of the dust grains \citep{Semenov+2003},
and the mass fraction of metals that condensed into dust grains.
We have assumed that all of the above properties follow the distributions or values found in the local ISM, which may not be applicable to the grains in the early universe.
Very likely, stellar dust sources (Asymptotic Giant branch stars, supernovae, pair-instability supernovae) 
and grain growth in the ISM provide different dust properties in low-metallicity environments 
compared to the present-day Universe \citep{Todini&Ferrara2001, Nozawa+2003, Schneider+2004, Ginolfi+2018}.
Since the size of the dust grain is smaller for dust produced by SNe, the cooling rate per unit mass is larger \citep{Schneider+2006},
while the total amount of dust grains can be one order of magnitude smaller due to destruction by the reverse shock 
\citep{Schneider+2012}.
Grain coagulation and grain growth via the depletion of the heavy element in the gas phase onto the grains during the cloud collapse increases the total mass of dust grains and significantly enhances the efficiency of dust cooling 
\citep{Hirashita&Omukai2009, Nozawa+2012, Chiaki+2015b}.
Considering all the above processes,
\citet{Chiaki+2016} have performed three-dimensional simulation and found that
when the gas phase metallicity is above $Z/Z_{\odot}=10^{-5}$,
dust cooling changes the thermal evolution and induces fragmentation,
which qualitatively agrees with our results 
assuming the standard-dust properties.
How the different dust properties and grain growth
affect the long term evolution and the stellar mass distribution is an important question and further studies are needed in the future.

%%%%%%%%%%%%%%%%%%%%%%%
%%%%%%%%%%%%%%%%%%%%%%%

%%%%%%%%%%%%%%%%%%%%%%%%
%%%%%%%%%%%%%%%%%%%%%%%%
\section{Summary}  
\label{sec::summary}
In this paper, we have followed the gravitational collapse of turbulent cloud cores
with initial metallicity in the range of $10^{-6} \lesssim Z/Z_{\odot} \lesssim 0.1$
and studied the impact of metallicity on the stellar mass distribution.
As noted by previous studies, we have found that the presence of dust grains 
promotes fragmentation and leads to the formation of 
low-mass stars with $0.01$--$0.1~M_{\odot}$ 
when the metallicity is larger than $10^{-5}~Z_{\odot}$.
There is a trend that the number of low-mass stars increases with metallicity.
Aside from the formation of low-mass stars,
the mass function is still top-heavy 
compared to the present-day IMF
in the low-metallicity environments with $Z/Z_{\odot} \lesssim 10^{-2}$.
In these cases, a massive stellar component appears at the cloud center.
These massive stars preferentially accrete the gas and efficiently grow in mass, resulting in the top-heavy mass distribution.
The mass distribution approaches the present-day IMF only for $Z/Z_{\odot} \simeq 10^{-1}$ case,
suggesting that the critical metallicity for the IMF transition is $Z/Z_{\odot}=0.01$--$0.1$.

Our results indicate that the stellar mass distribution becomes top-heavy
when turbulent motion cannot grow 
due to inefficient cooling in the low-density regime $n\sim 10^{6}~\mathrm{cm^{-3}}$.
In this case, star formation begins only after the turbulence decays,
as found when $Z/Z_{\odot} \lesssim 0.01$.
Subsequently, a single massive core monolithically collapses.
The central stellar system is formed surrounded by a massive gas disk, which efficiently feeds the central massive star or drives low-mass stars to migrate inward.
This makes the mass distribution more top-heavy than the present-day IMF.

When $Z/Z_{\odot} \simeq 10^{-1}$,
fine-structure line cooling is efficient
at the scales where turbulent motion dominates,
leading to the formation of a filamentary structure.
If the filament becomes dense enough for gravitational instability to operate,
it fragments into protostellar cores.
In this case, the size of the circumstellar disks around massive stars is 
much smaller than in $Z/Z_{\odot} \lesssim 0.01$ cases, since the accreting gas has different orientation of the angular momentum 
owing to turbulent motion,
preventing the formation of a large disk.
As a result, the mass supply rate to massive stars becomes smaller than in the monolithic collapse cases.

Our results suggest that the properties of turbulence seeded in the initial cores is crucial for setting the mass distribution.
Such turbulence is mainly driven by physical processes in the external environment,
i.e. the expansion of H~{\sc ii} shells or 
the injection of a large amount of kinetic energy by SN explosions.
How the turbulence properties influence the mass distribution, as well as its interplay between the cooling and heating due to the detailed radiative processes should be further studied.

%%%%%%%%%%%%%%%%%%%%%%%%
%%%%%%%%%%%%%%%%%%%%%%%%
\section*{Acknowledgements}

This work is financially supported by
the Grants-in-Aid for Basic Research by the Ministry of Education, Science and Culture of Japan 
(SC:19J00324, KO:25287040, 17H01102, 17H02869). 
RS acknowledges support from the Amaldi Research Center funded by the MIUR program Dipartimento di Eccellenza (CUP:B81I18001170001) and funding from the INFN TEONGRAV specific initiative.
We conduct numerical simulation on XC50 at the Center for Computational Astrophysics (CfCA) of the National Astronomical Observatory of Japan
and XC40.
We also carry out calculations on XC40 at YITP in Kyoto University.
The work was also conducted using the resource of Fujitsu PRIMERGY CX2550M5/CX2560M5(Oakbridge-CX) 
in the Information Technology Center, The University of Tokyo.
We use the SPH visualization tool SPLASH \citep{SPLASH} in Figs.~\ref{fig_snapshot} and \ref{fig_snapshot_m150}.

\section*{DATA AVAILABILITY}
The data underlying this article will be shared on reasonable request to the corresponding author.

\bibliography{biblio2}

\end{document}